\documentclass[prd,superscriptaddress,showpacs]{revtex4}

\usepackage{epsfig}

\newcommand{\tfrac}[2]{{\textstyle{\frac{#1}{#2}}}}
\newcommand{\ep}{\varepsilon}
\newcommand{\OS}{\text{OS}}
\newcommand{\msb}{\overline{\text{MS}}}
\newcommand{\MSb}{$\overline{\text{MS}}$ }

\begin{document}

\title{Bosonic Corrections to $\Delta r$ at the Two Loop Level}

\author{M. Awramik}

\author{M. Czakon}

\affiliation{Institut f\"ur Theoretische Physik, 
Universit\"at Karlsruhe, D-76128 Karlsruhe, Germany}

\affiliation{Department of Field Theory and Particle Physics,
Insitute of Physics, University of Silesia,
Uniwersytecka 4, PL-40007 Katowice, Poland}

\author{A. Onishchenko}

\affiliation{State Research Center, Institute for High Energy Physics,
Protvino, Moscow Region, 142284 Russia}

\affiliation{Institut f\"ur Theoretische Teilchenphysik, 
Universit\"at Karlsruhe, D-76128 Karlsruhe, Germany}

\author{O. Veretin}

\affiliation{Institut f\"ur Theoretische Teilchenphysik, 
Universit\"at Karlsruhe, D-76128 Karlsruhe, Germany}

\begin{abstract}
The details of the recent calculation of the two-loop bosonic
corrections to the muon lifetime in the Standard Model are
presented. The matching on the Fermi theory is
discussed. Renormalisation in the on-shell and in the \MSb scheme is
studied and transition between the schemes is shown to lead to
identical results. High precision numerical methods are compared with
mass difference and large mass expansions.
\end{abstract}


\maketitle


\section{Introduction}

The muon decay lifetime ($\tau_\mu$) has been used for long as an
input parameter for high precision predictions of the Standard Model
(SM). It allows for an indirect determination of the mass of the $W$
boson ($M_W$), which suffers currently from a large experimental error
of 39~MeV \cite{Hagiwara:pw}, one order of magnitude worse than that
of the $Z$ boson mass ($M_Z$). A reduction of this error by LHC to
15~MeV \cite{lhctdr} and by a future linear collider to 6~MeV
\cite{tesla-tdr} would provide a stringent test of the SM by
confronting the theoretical prediction with the experimental value.

The extraction of $M_W$ with an accuracy matching that of next
experiments, {\it i.e.} at the level of a few MeV necessitates
radiative corrections beyond one loop order. Large two-loop
contributions from fermionic loops have been calculated in
\cite{Freitas:2000gg}.  The current prediction is affected by two
types of uncertainties. First, apart from the still unknown Higgs
boson mass, two input parameters introduce large errors. The current
knowledge of the top quark mass results in an error of about 30~MeV
\cite{Freitas:2002ja}, which should be reduced by LHC to 10~MeV and by
a linear collider even down to 1.2~MeV. The inaccuracy of the
knowledge of the running of the fine structure constant up to the
$M_Z$ scale, $\Delta \alpha(M_Z)$, introduces a further $6.5$~MeV
error. Second, several higher order corrections are unknown. In fact
the last unknown correction at the ${\cal O}(\alpha^2)$ order has been
calculated only recently in \cite{my} and \cite{oni}. This contribution
comes from diagrams with no closed fermion loops.

It is the purpose of the present work to give a detailed description
of the methods used in the calculations presented in \cite{my} and
\cite{oni}. Since one of the groups used high precision numeric
methods and the other deep expansions both in mass differences and in
large masses, a comparison can be given.

In the next section we discuss the question of matching of the Fermi
theory onto the Standard Model at low energy scales. Then we move to
the discussion of renormalisation in the on-shell scheme and continue
with the \MSb scheme. A section on the transition between the schemes
contains comparisons of the methods and the final results. The
description of computational methods and conclusions close the main
part of the work. In the appendices, a derivation of the electric
charge counterterm through the $U(1)$ Ward identity can found,
followed by the explicit analytic results of the expansions of the
on-shell and the \MSb quantities.


\section{Matching}

\label{matching}

The muon lifetime $\tau_\mu$ can be computed from the effective
Fermi theory given by the lagrangian
\begin{equation}
  \label{Fermilagr}
  {\cal L}^{\rm eff} = {\cal L}_{\rm QED} + 
  \frac{G_F}{\sqrt2}\,O_F + \mbox{higher dimension operators} \,,
\end{equation}
where $O_F$ is the 4-fermion Fermi operator of dimension six
\begin{equation}
  O_F =  \bigl[\bar{\nu}_\mu \gamma^\alpha (1-\gamma_5) \mu \bigr]
  \,   \bigl[  \bar{e} \gamma_\alpha (1-\gamma_5) \nu_e \bigr] ,
  \label{fermiop}
\end{equation}
and $G_F$ is the Fermi constant. Note that Eq. \ref{Fermilagr} is
a definition of $G_F$.  This lagrangian can be used to describe low
energy processes (such that energies are $\ll M_W$) mediated by the
weak charged current. Since the theory Eq.~\ref{Fermilagr} is
nonrenormalisable, an ultraviolet cut-off $\Lambda$ should be
introduced.

In particular for the muon decay process we have
\begin{equation}
  \frac{1}{\tau_\mu} = \frac{G_F^2 m_\mu^5}{192\pi^3} 
  \left( 1 - 8\frac{m_e^2}{m_\mu^2}\right) (1+\Delta q) \,,
\end{equation}
with $m_e$ and $m_\mu$ being the masses of the electron and the muon
respectively. The quantity $\Delta q$ describes all QED corrections
in the Fermi theory and has been calculated at the one-loop
\cite{Berman:1958ti} and at the two-loop \cite{vanRitbergen:1998yd}
order.

By its nature $G_F$ is the Wilson coefficient function of the operator
$O_F$ and can be evaluated from the SM. Traditionally the matching
relation between $G_F$ and the parameters of the SM is parametrised as
follows
\begin{equation}
  \frac{G_F}{\sqrt2} = \frac{e^2}{8(1-M_W^2/M_Z^2)M_W^2}(1+\Delta r).
  \label{drdefinition}
\end{equation}
The quantity $\Delta r=\Delta r^{(1)}+\Delta r^{(2)}+\dots$ absorbs
the effects of all loop diagrams.

It is the purpose of the following subsection to establish the
framework for the calculation of $\Delta r$. 

\subsection{Factorisation theorem}

In principle, the muon decay amplitude can be evaluated directly in
the SM, but this is not feasible in practice. There are many scales
involved which vary from less than 1 MeV to 100 GeV, {\it i.e.} by
more than 5 orders of magnitude! On the other hand the number of
Feynman diagrams grows very fast with the number of loops. A way out
to keep the problem manageable is to switch on the machinery of
effective lagrangians (see Eq.~\ref{Fermilagr}). This allows one to
simplify the calculation enormously and to separate consistently the
low energy (``soft'') dynamics from high energy (``hard'') static
characteristics.

Suppose that we can compute the muon decay amplitude $A^{\text{SM}}$
in the SM.  Then the Fermi constant $G_F$ defined through
Eq.~\ref{Fermilagr} can be predicted from $A^{\text{SM}}$. Indeed we
should require that both evaluations in the SM and the Fermi theory
give the same result. At the tree level the corresponding matching
equation reads
\begin{eqnarray}
\label{mateq1}
  A^{\text{SM}} = \frac{G_F}{\sqrt2} \,
  \langle \mu| O_F | e\nu_\mu\bar{\nu_e} \rangle 
  + {\cal O}\left(\frac{m_\mu^4}{M_W^4}\right)  \,.
\end{eqnarray}
This equation just states that the amplitude of the process $\mu\to
e\nu\bar{\nu}$ is the same both in the full SM and in the effective
Fermi theory up to operators of higher dimension.

When loop effects are taken into account, matrix elements in both
sides of Eq.~\ref{mateq1} get quantum corrections.  Since
$A^{\text{SM}}$ and $\langle \mu| O_F | e\nu_\mu\bar{\nu_e} \rangle$
are amputated matrix elements one has to renormalise also the external
wave functions. Therefore the final form of the matching equation reads
\begin{eqnarray}
  \label{mateq2}
  \sqrt{ Z_{2,e}^{\text{SM}} Z_{2,\mu}^{\text{SM}} 
    Z_{2,\nu_e}^{\text{SM}} Z_{2,\nu_\mu}^{\text{SM}} } \,\,
  A^{\text{SM}} =
  \sqrt{ Z_{2,e}^{\text{eff}} Z_{2,\mu}^{\text{eff}} 
    Z_{2,\nu_e}^{\text{eff}} Z_{2,\nu_\mu}^{\text{eff}} }  \,
  Z_{O_F}^{-1} \,\,
  \frac{G_F}{\sqrt2} \,
  \langle \mu | O_F | e\nu_\mu\bar{\nu_e} \rangle 
  + {\cal O}\left(\frac{m_\mu^4}{M_W^4}\right) \, ,
\end{eqnarray}
where $Z_{2,f}^{\text{SM}}$ and $Z_{2,f}^{\text{eff}}$ are
wave function renormalisation constants of the fermions
evaluated in the SM and in the effective theory respectively and
$Z_{O_F}$ is the renormalisation constant of the Fermi operator
in the effective theory.

There are two ways to compute $G_F$ from the SM:
\begin{enumerate}
\item standard matching calculation, or
\item automatic matching via factorisation theorem. 
\end{enumerate}

The former approach works always by simply computing all ingredients
(apart from $G_F$) in the matching equation Eq.~\ref{mateq2}. This
requires however much extra efforts to evaluate the ``soft'' pieces
(or, at least, to separate them) in the amplitudes and the $Z$'s.
Historically for this purpose the Pauli--Villars regularisation was
used in \cite{Sirlin1} and then extended to two-loop order in
\cite{Sirlin2}. The same approach has been applied also in
\cite{Freitas:2002ja}.

\begin{figure}
\epsfig{file=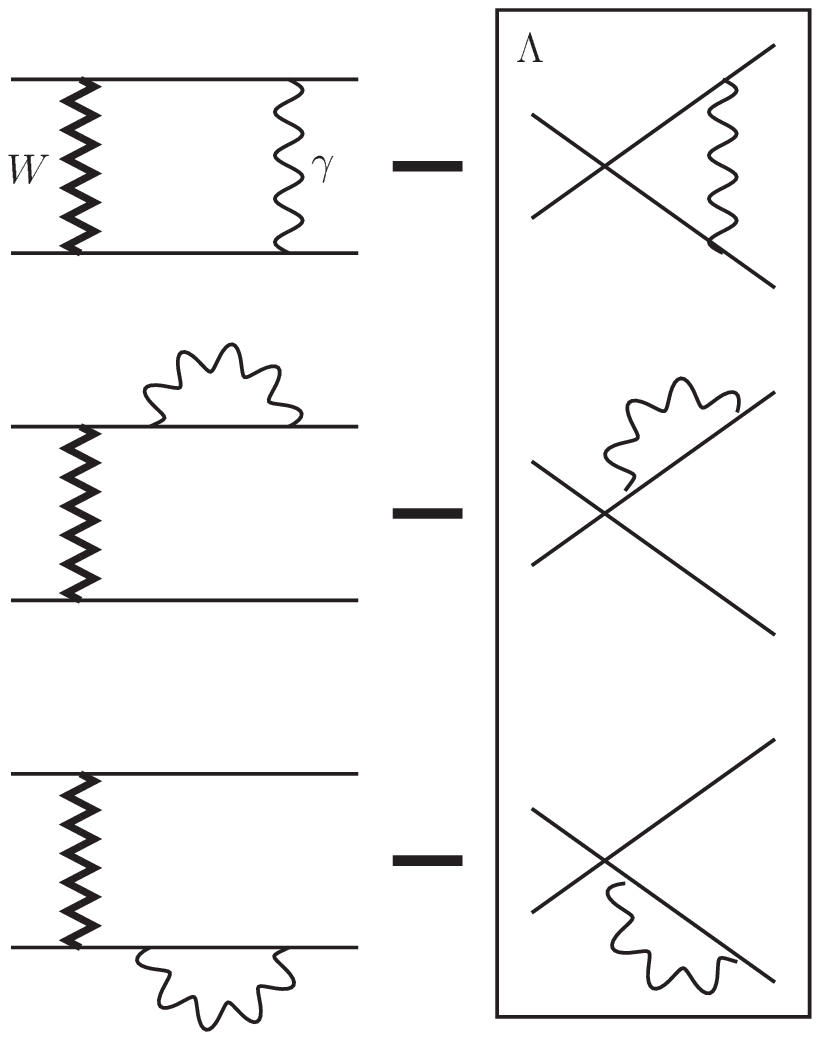, width=6cm}
\caption{\label{factorisation}
one-loop factorisation with Pauli-Villars regularisation.}
\end{figure}

How it works at the 1-loop level is demonstrated in
Fig.~\ref{factorisation}. There are only three infrared divergent
diagrams with photon. From each diagram its counterpart in the Fermi
theory should be subtracted.  The left diagram in each line of
Fig.~\ref{factorisation} corresponds to the  result in the full model
and therefore contains both the ``soft'' and the ``hard'' part. The
right one contains only the ``soft'' part, which means that the
difference is the requested ``hard'' correction. In addition, for the
diagrams in the frame the Pauli-Villars regularisation is introduced
to regularise the ultraviolet divergences.  At the two-loop level we
have a very similar situation.  The difference is that instead of
``hard'' and ``soft`' terms there are now ``hard-hard'',
``hard-soft'', ``soft-hard'' and ``soft-soft'' contributions.  From
these only the ``hard-hard'' piece contributes to $G_F$.

Accidentally, it happens that the sum of the three ``soft'' diagrams
inside the frame in Fig.~\ref{factorisation} is an ultraviolet finite
quantity (let us call it $\Sigma_{\text{soft}}$).  It is easy to prove
that this holds true also to all orders.  This is a consequence of the
Ward--Takahashi identity for $QED$.  This fact, however, is a pure
coincidence rather than something fundamental. If such a cancelation
had not occurred, renormalisation of the operator $O_F$ would be
required as it is taken into account in Eq. \ref{mateq2}.

The scheme given in Fig.~\ref{factorisation} is consistent but the
disadvantage of it  is that there arises the problem  of bookkeeping
of ``soft'' and ``hard'' parts and already at  the two-loop level the
problem becomes very complicated. Indeed, at the two-loop level one
has to subtract from each diagram the ``hard-soft'', ``soft-hard'' and
``soft-soft'' pieces.

Therefore it would be very helpful to find some other way to obtain
the ``hard'' part.  Thus we come to the second way to compute
$G_F$---automatic matching.  This procedure is the most
straightforward and the most economical  (minimal in costs) way to
compute. It is based on the factorisation theorem, proven e.g. in
\cite{Gorishnii}. It allows one to extract the ``hard'' part directly
without any reference to ``soft'' pieces.  As a well known example of
such a procedure we can mention the  evaluation of Wilson coefficient
functions in deep inelastic scattering processes.

Returning to the sum of the three ``soft'' graphs in
Fig.~\ref{factorisation} ($\Sigma_{\text{soft}}$) we notice that in
$G_F$ all ``soft'' modes are eliminated.  This means, that all
subgraphs in Fig.~\ref{factorisation} should be computed at vanishing
masses of the leptons. In this case the Ward--Takahashi identity not
only makes $\Sigma_{\text{soft}}$ ultraviolet finite but also nullifies
it.  Thus all ``soft'' parts add up to zero. This is also true to all
orders of perturbation theory.  In other words, one can from the very
beginning  nullify all external momenta and masses and evaluate the
obtained bubble diagrams.  Of course new infrared divergences are
generated.  They cancel however in the expression for $G_F$.  To
regularise these infrared divergences we use the dimensional
regularisation.
   
To prove rigorously that infrared singularities indeed drop out from
the result one can turn to the framework for construction of effective
low energy lagrangians given in \cite{Gorishnii}.  At the level of
individual Feynman diagrams one can separate ``soft'' and ``hard''
scales with the help of the asymptotic expansion  procedure
\cite{asymptotic}.  Let $F$ denote a Feynman diagram. Then
\begin{equation}
  \label{asym}
  F \sim \sum\limits_{H \subseteq F} S \cdot T(H)  \,,
\end{equation}
where the sum runs over all ``hard'' subgraphs $H$ of the diagram $F$;
$S$ is a ``soft'' subgraph obtained from $F$ by shrinking $H$ to a
point  and $T$ stands for the Taylor expansion (before integration!)
of $H$ with respect to all ``soft'' parameters.  The exact rules for
construction of hard subgraphs are discussed in details in
\cite{asymptotic}.

The important property of the operation Eq.~\ref{asym} is that it has
the combinatorial structure of the $R$-operation
\cite{Roperation}. This allows one to promote the operation on a
single Feynman diagram to the operation on the whole Feynman amplitude
(the factorisation theorem).  By this procedure all infrared
divergencies are absorbed either  by the ``soft'' matrix element or by
the renormalisation constant $Z_O$ of the operator. The detailed
discussion can be found in \cite{Gorishnii}.

In the case of $G_F$ we have further simplifications.
\begin{itemize}
\item The anomalous dimension of the Fermi operator ${O_F}$ is zero,
therefore $Z_{O_F}$ in the matching equation Eq.~\ref{mateq2} is equal
to one.
\item At zero lepton masses and external momenta all
$Z_2^{\text{eff}}$ and the ``soft'' matrix element in Eq.~\ref{mateq2}
are equal to one.
\end{itemize}

Finally we get
\begin{eqnarray}
  \label{mateq3}
  \frac{G_F}{\sqrt2} = \Biggl[
  \sqrt{ Z_{2,e}^{\text{SM}} Z_{2,\mu}^{\text{SM}} 
    Z_{2,\nu_e}^{\text{SM}} Z_{2,\nu_\mu}^{\text{SM}} } \,\, 
  A^{\text{SM}} \Biggr]_{\rm hard}\,,
\end{eqnarray}  
where the subscript ``hard'' means that all ``soft'' scales
are put to zero.

Thus the problem is reduced completely to the vacuum Feynman diagrams
of one- and two-loop order and the bookkeeping problem does not arise
at all.  The wave function renormalisation constants are to be
computed in the on-shell scheme. Again, for massless leptons, the wave
function renormalisation constants are defined through vacuum diagrams
only.  Such diagrams can be evaluated analytically using reduction
formulae of \cite{Davydychev:1992mt} based on integration by parts
identities \cite{Chetyrkin:qh}.

\subsection{Projection}

An important problem in the calculation is the reduction of the
amplitudes to scalar integrals. It is not only of practical
importance. In fact it is connected to the correct definition of the
matrix elements in the model, since dimensional regularisation is
used.

The matching onto the Fermi theory with its double $V-A$ chiral
structure is made possible because of the left-handedness of the
charged current in the Standard Model. The ``hard'' components of the
diagrams contain only massless fermions and therefore formally the
structure of the two spinor lines can be mapped onto the operator
\begin{equation}
  \label{v-a}
  \gamma^\mu P_L \otimes \gamma_\mu P_L.
\end{equation}
In four dimensions, every string of an odd number of gamma matrices
and a left-handed projector can be reduced to the structure
$\gamma^\mu P_L$ due to the Chisholm identity
\begin{equation}
  \gamma_\mu \gamma_\nu \gamma_\rho = g_{\mu\nu} \gamma_\rho +
  g_{\nu\rho} \gamma_\mu - g_{\mu\rho} \gamma_\nu -i
  \epsilon_{\mu\nu\rho\sigma} \gamma^\sigma \gamma_5.
\end{equation}
The reduction leads to the operator
\begin{equation}
  T_{\mu\nu} \gamma^\mu P_L \otimes \gamma^\nu P_L,
\end{equation}
where $T_{\mu\nu}$ is some tensor made of the integration
momenta. Since there are no non-vanishing external momenta, this tensor
must be proportional to $g_{\mu\nu}$ and the result Eq.~\ref{v-a}
follows. A suitable way to obtain directly the right value is to use a
projector made of trace operators. Let the original product of strings
of gamma matrices be denoted by
\begin{equation}
  \Gamma_1 \otimes \Gamma_2.
\end{equation}
We wish to obtain the proportionality coefficient $A$ in the following
equation
\begin{equation}
  \int \Gamma_1 \otimes \Gamma_2 = A \times (\gamma^\mu P_L \otimes
  \gamma_\mu P_L ).
\end{equation}

\begin{figure}
\epsfig{file=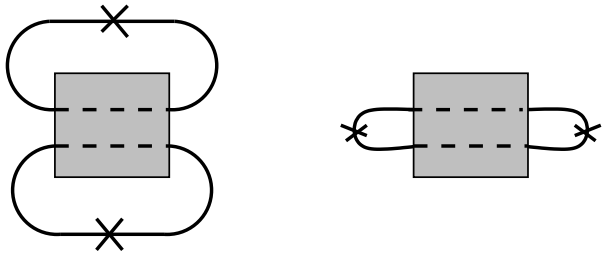,width=8cm}
\caption{\label{projector}
Two of the possible projectors for $\Delta r$. The dashed lines
represent the strings of Dirac matrices, while the crosses, the
projection operators.}
\end{figure}

Two possibilities of closing the spinor strings with trace operators
are depicted in Fig.~\ref{projector}. The left one has been used in
\cite{Freitas:2002ja} and is given by the equation
\begin{equation}
  \label{proj1}
  A = \frac{1}{4 d} \int \text{Tr} (\Gamma_1 \gamma^\mu P_R)
  \text{Tr} (\Gamma_2 \gamma_\mu P_R),
\end{equation}
where the dimension of space-time $d$ has been kept arbitrary and the
trace of the unit matrix has been put to 4, as usual. A second
possibility which we used to perform the calculations presented in
this work is given by
\begin{equation}
  \label{proj2}
  A = -\frac{1}{2 d (d-2)} \int \text{Tr} (\Gamma_1 \gamma_\mu P_R \Gamma_2
  \gamma^\mu P_R),
\end{equation}
and corresponds to the right picture in Fig.~\ref{projector}

Both projectors are obviously equivalent in four dimensions due to the
Chisholm identity as explained above. The difference starts to be
important for divergent integrals. In fact the problem does only occur
for one-particle-irreducible four-point diagrams, where the divergence
can come from two sources. First from the external wave function
renormalisation, which is incomplete due to infrared divergences and
second due to infrared divergences of the diagrams themselves. As
noticed in \cite{Freitas:2002ja} the first projector Eq.~\ref{proj1}
needs to be corrected, as it does not fulfil several requirements,
like for example the vanishing of diagrams with propagator insertion
in the photon lines. Moreover, one can explicitely check that without
corrections the subtracted diagrams in the Pauli--Villars approach do
not cancel and the dependence on the $\Lambda$ scale remains. In the
automatic factorisation approach this shows up through an incomplete
cancellation of divergences. Notice, however that the result is gauge
independent, thus it is only the finiteness of the result that shows
that the projector is incorrect.

On the contrary the projector Eq.~\ref{proj2} does not require any
corrections. It does fulfil all of the algebraic requirements and
also yields a finite result as well as the exact cancellation of the
subtraction diagrams of Fig.~\ref{factorisation} in $d$-dimensions and
in all orders of perturbation theory. This useful property follows
from the fact that this projector respects the Fierz symmetry in
$d$-dimensions. One can check explicitely that for example
\begin{equation}
  \gamma_\mu \gamma_\nu \gamma_\rho P_L \otimes \gamma^\rho \gamma^\nu
  \gamma^\mu P_L \sim \gamma_\mu \gamma_\nu \gamma_\rho \gamma^\nu
  \gamma^\mu P_L \otimes \gamma^\rho P_L,
\end{equation}
where $\sim$ means equality after projection.


\section{On-shell renormalisation}

Two-loop calculations within the on-shell renormalisation scheme
require the knowledge of several counterterms. At the very least
charge and mass counterterms are needed. In this section we first
discuss the problem of gauge invariance in connection with tadpole
diagrams. We then give specific expressions for the required
counterterms.

\subsection{Tadpoles and gauge invariance of counterterms}

\label{tadpoles}

It has been known for a long time 
that the inclusion of tadpoles is necessary to
obtain gauge invariant counterterms. In fact this property has been
first noticed \cite{Appelquist:ms} shortly after the proof of
renormalisability of gauge theories. A general proof of the Quantum
Action Principle, which has for consequence the gauge invariance of
on-shell processes in the bare lagrangian, requires the inclusion of
even those tadpoles which would be cancelled by normal ordering (one
loop tadpoles) \cite{Breitenlohner:hr}. There are, however, two
disadvantages of using tadpoles in actual calculations. First, this
requires the inclusion of diagrams, which drop in the final
result. Second, one-particle-irreducible (1PI) Green functions cannot
contain tadpoles. As long as we wish to obtain results at the least
cost and by using automated software, it is interesting to consider
alternative possibilities.

It turns out that it is possible to prepare the bare lagrangian in
such a way, that the only gauge dependent quantities would be the wave
function renormalisation constants and the vacuum renormalisation
constant, and still all of the tadpoles would be cancelled. Let us
start by considering a lagrangian in which the bare coupling and
masses are defined through physical processes. The masses can be
equivalently defined through the position of the poles of the physical
$S$-matrix in the complex plane as recently proved
\cite{Gambino:1999ai}. In such a case all of the bare parameters would
be gauge invariant, because they would fulfil equations that have this
same property. It is important to supply a condition on the vacuum
expectation value of the bare Higgs field $v_0$ that would resum terms
of order ${\cal O}(\alpha^0)$. A choice which is still consistent with
gauge invariance is
\begin{equation}
  \label{tadp}
  \frac{1}{2} v_0 \left( \frac{1}{2} v_0^2 \lambda_0-\mu_0^2 \right) = 0,
\end{equation}
where $\lambda_0$ and $\mu_0$ are defined through the Higgs
lagrangian
\begin{equation}
  {\cal L}_{\text{Higgs}} = \frac{1}{2} \mu_0^2 \Phi_0^\dagger \Phi_0
  -\frac{1}{4} \lambda_0 (\Phi_0^\dagger \Phi_0)^2,
\end{equation}
and $\Phi_0$ is the Higgs doublet. Eq.~\ref{tadp} implies the
vanishing of the linear term in the lagrangian. Although this term
will be subsequently altered, the tree level contribution will always
vanish.

\begin{figure}
\epsfig{file=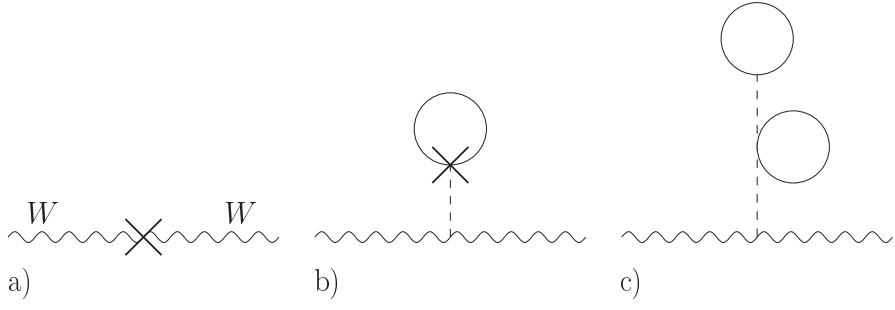, width=12cm}
\caption{\label{tadpCT}
Vacuum expectation value counterterm insertion into the $W$ boson
self-energy a), reproducing tadpole insertions b) and c).}
\end{figure}

We now introduce an additional  renormalisation of the bare
vacuum expectation value
\begin{equation}
  v_0 \longrightarrow v_0 Z_v^{1/2}.
\end{equation}
The renormalisation constant $Z_v$ can be used to cancel
the tadpoles recursively, which implies together with Eq.~\ref{tadp} that the
first non-vanishing term in its perturbative expansion starts at order
${\cal O}(\alpha)$. The linear term in the Higgs field can now be
written as
\begin{equation}
  -T_0 H^0 = -\frac{M^0_W \sin \theta_W^0}{e_0} (M^0_H)^2 Z_v^{1/2}
  (Z_v-1) H^0
\end{equation}
where the following relations have been used
\begin{eqnarray}
  v_0 &=& \frac{2 \sin \theta_W^0 M_W^0}{e_0}, \\
  \mu^2_0 &=& (M_H^0)^2, \\
  \lambda_0 &=& \left( \frac{e_0 M_H^0}{\sin \theta_W^0 M_W^0} \right)^2.
\end{eqnarray}
At the tree level the contribution is zero, since then $Z_v^{(0)} = 0$,
as noticed above.  To one-loop order, the relation between the tadpole
diagrams and the vacuum expectation value is simple
\begin{equation}
  \delta Z_v^{(1)} = \frac{e}{\sin \theta_W M_W M_H^2} \Pi_H^{(1)},
\end{equation}
where $i \Pi_H^{(1)}$ is the sum of 1PI one-loop tadpole diagrams of the
Higgs field. The situation gets much more complicated at the two-loop
level
\begin{equation}
  \label{twotadp}
  \delta Z_v^{(2)} = \frac{e}{\sin \theta_W M_W M_H^2} \Pi_H^{(2)}
  - \frac{1}{2} \delta Z_v^{(1)} \left( \delta Z_v^{(1)}
  + \delta Z_H^{(1)} + 2 \frac{\delta M_H^{2(1)}}{M_H^2}
  + \frac{\delta M_W^{2(1)}}{M_W^2} + 2 \frac{\delta \sin 
      \theta_W^{(1)}}{\sin \theta_W} - \delta Z_e^{(1)} \right).
\end{equation}
At this level an insertion of this counterterm reproduces all of the
tadpole diagrams that would be included in the usual approach. An
example is depicted in Fig.~\ref{tadpCT}. An insertion of $Z_v$ into
the $W$ boson self-energy a), leads effectively through the first term in
Eq.~\ref{twotadp} to an insertion of a one-loop tadpole with with a
vertex counterterm b). This counterterm also contains a correction to
the Vacuum expectation value of the Higgs field, which reproduces the
tadpole diagram c).

\subsection{On-shell scheme counterterms}

\label{ctos}

The on-shell renormalisation scheme is defined by the requirement that
the masses be identified through the poles of the physical $S$-matrix
(as the real part of the pole), while the electric charge coincide
with the value measured in the Thompson scattering process as for
example in the quantum Hall effect. These conditions are enough to fix
all of the free parameters of the SM with minimal Higgs sector
(neglecting the CKM matrix and the strong coupling constant). The
counterterms have been given by many authors. The peculiarity of the
present work is the specific definition of the bare masses which are
gauge invariant without including tadpole diagrams. This, however,
implies that the formulae defining the counterterms will be slightly
different.

At the one-loop level, the mass counterterms are related
to the on-shell self-energies through
\begin{eqnarray}
  \delta M_H^{2(1)} &=& {\text{Re}}
  \Pi_{HH}^{(1)}(M_H^2)-\frac{3}{2}M_H^2 \delta Z_v^{(1)}, \\
  \delta M_W^{2(1)} &=& -\Pi_{WW,T}^{(1)}(M_W^2)-M_W^2 \delta Z_v^{(1)}, \\
  \delta M_Z^{2(1)} &=& -\Pi_{ZZ,T}^{(1)}(M_Z^2)-M_Z^2 \delta Z_v^{(1)},
\end{eqnarray}
where $i \Pi_{ii,T}$ denotes the transverse part of the self-energy
diagrams of the boson $i$. For bosonic corrections to the Higgs
boson mass counterterm the real part has to be taken due to the
possible decay into a $W$ or $Z$ boson pair. To one-loop order this
still yields a gauge invariant result for the renormalised
amplitude. The $W$ and $Z$ bosons do not require such a treatment
neither at one nor at two-loop order.

At the two-loop order, only $W$ and
$Z$ boson mass counterterms are needed, and they assume the form
\begin{eqnarray}
  \delta M_W^{2(2)} &=& -\Pi_{WW,T}^{(2)}(M_W^2)-\delta Z_W^{(1)} 
  \delta M_W^{2(1)} -M_W^2 \delta Z_v^{(2)} - \delta Z_v^{(1)} 
  (M_W^2 \delta Z_W^{(1)}+\delta M_W^{2(1)}), \\
  \delta M_Z^{2(2)} &=& -\Pi_{ZZ,T}^{(2)}(M_Z^2)-\delta Z_Z^{(1)} 
  \delta M_Z^{2(1)} - M_Z^2 \delta Z_v^{(2)} - \delta Z_v^{(1)} 
  (M_Z^2 \delta Z_Z^{(1)}+\delta M_Z^{2(1)}) +\frac{1}{4} M_Z^2 (\delta
  Z_{\gamma Z}^{(1)})^2.
\end{eqnarray}
The last term in the $Z$ boson mass counterterm, which does not occur
in the $W$ boson mass counterterm, has its origin in the mixing
between $Z$ and $\gamma$. If the self-energies have imaginary parts,
then suitable additional terms have to be included as described in
\cite{Freitas:2002ja}. The above formulae are valid only if the
subdivergencies in the two-loop self-energies are renormalised.  They
also require the wave function renormalisation constants of the bosons
\begin{eqnarray}
  \delta Z_W^{(1)} &=& \Pi^{(1)'}_{WW,T} (M_W^2), \\
  \delta Z_Z^{(1)} &=& \Pi^{(1)'}_{ZZ,T} (M_Z^2),
\end{eqnarray}
and the mixing renormalisation
\begin{equation}
  \delta Z_{\gamma Z}^{(1)} = \frac{2}{M_Z^2} \Pi_{\gamma Z,T}^{(1)} (M_Z^2).
\end{equation}
The last two constants form part of the $2 \times 2$ renormalisation
matrix of the neutral bosons
\begin{equation}
  \left(
    \begin{array}{c} A^0_\mu \\ Z^0_\mu \end{array} 
  \right) = 
  \left(
    \begin{array}{cc} Z^{1/2}_{ZZ} & \frac{1}{2} Z_{\gamma Z} \\
      \frac{1}{2} Z_{Z \gamma} & Z^{1/2}_{\gamma \gamma} \end{array}
  \right)
  \left(
    \begin{array}{c} A_\mu \\ Z_\mu \end{array} 
  \right) .
\end{equation}
The remaining two renormalisation constants define the photon field
and can be obtained at zero momentum transfer from the following
formulae
\begin{eqnarray}
  \delta Z^{(1)}_{Z \gamma} &=& -\frac{2}{M_Z^2} \Pi^{(1)}_{\gamma
  Z,T}(0), \\
  \delta Z^{(1)}_{\gamma \gamma} &=& \Pi^{(1)'}_{\gamma \gamma,T} (0).
\end{eqnarray}

The electric charge counterterm can be obtained in two ways. The first
consists in simply calculating the scattering of fermions off real
photons, {\it i.e.} at zero momentum transfer. This however introduces
unnecessarily three point functions. A second possibility is to use the
$U(1)$ Ward identity. The suitable relation between the wave function
renormalisation constants of the photon and the $Z$ boson has been
proved in \cite{bauberger:phd} using the $BRS$ symmetry. A simpler
proof is given in appendix~\ref{ward}. The one and two-loop
counterterms in the on-shell scheme are given by
\begin{eqnarray}
  \delta Z_e^{(1)} &=& -\frac{1}{2} \delta Z^{(1)}_{\gamma\gamma} -
  \frac{1}{2} \frac{\sin \theta_W}{\cos \theta_W} \delta
  Z^{(1)}_{Z\gamma}, \\
  \delta Z_e^{(2)} &=& -\frac{1}{2} \delta Z^{(2)}_{\gamma\gamma} -
  \frac{1}{2} \frac{\sin \theta_W}{\cos \theta_W} \delta
  Z^{(2)}_{Z\gamma} + (\delta Z_e^{(1)})^2+\frac{1}{8} (\delta 
  Z^{(1)}_{\gamma\gamma})^2-\frac{1}{2} \frac{\delta \sin
  \theta_W}{\cos^3 \theta_W} \delta Z^{(1)}_{Z \gamma}.
\end{eqnarray}
The two-loop wave function renormalisation of the photon is given by
the short formula
\begin{equation}
  \delta Z^{(2)}_{\gamma\gamma} = \Pi^{(2)'}_{\gamma\gamma,T}(0) -
  \frac{1}{4} (\delta Z^{(1)}_{Z\gamma})^2,
\end{equation}
whereas in the mixing counterterm, the vacuum expectation value
correction makes again its appearance
\begin{equation}
  \delta Z^{(2)}_{Z\gamma} = -\frac{2}{M_Z^2} \Pi^{(2)}_{\gamma Z}(0)
  - \frac{1}{2} \delta Z^{(1)}_{ZZ} \delta Z^{(1)}_{Z\gamma}
  - \frac{1}{M_Z^2} \delta Z^{(1)}_{Z\gamma} \delta M_Z^{2(1)}
  - \delta Z^{(1)}_v \delta Z^{(1)}_{Z\gamma}.
\end{equation}
In the on-shell calculation the ghost sector was also
renormalised. The respective constants are as in \cite{Freitas:2002ja}
up to an unimportant renormalisation of the ghost wave functions,
the difference being dictated by simplicity. The wave function
renormalisation constants of the ghosts and Goldstone bosons have been
left unspecified. For the ghosts, these constants cancel trivially
within every closed loop. With the Goldstone bosons, the situation is
more complicated, since the fact that the gauge fixing term should not
be renormalised induces Goldstone wave function renormalisation
constants in the ghost sector. These can only cancel in gauge
invariant quantities. This indeed happened for all the mass and
coupling counterterms and for the complete result.


\section{\MSb renormalisation}

\label{msrenor}

In this section we describe in detail the renormalisation of $\Delta
r$ in the \MSb scheme. $\Delta r$ is computed through the matching
procedure described before in Section~\ref{matching}.  Here we chose
the strategy of multiplicative renormalisation.  After multiplication
by the on-shell wave function renormalisation constants of external
fermion fields, the result is expressed in terms of bare masses and
bare electric charge.  In order to get the \MSb renormalised result
for $G_F$ one needs to substitute all bare parameters in the form
\begin{eqnarray}
  e_0 &=& \mu^\ep \, Z_e \, e(\mu) , \nonumber\\
  (m_i^0)^2 &=& Z_{m_i} m^2_i(\mu) ,
  \label{MSbarren} 
\end{eqnarray}
where $e(\mu)$ and $m_i(\mu)$ are the \MSb charge and masses
respectively and $\mu$ is the \MSb parameter. The \MSb renormalisation
constants will be specified in the next two subsections.

Let us stress that in Eq. \ref{MSbarren} we renormalise only  the
physical parameters and no renormalisation of the unphysical sector
(ghost sector and gauge fixing parameters) is required. The
renormalisation of the boson particles' wave functions is also not
needed since it cancels anyway in the final expression.

A few words should also be said about tadpole diagrams, which should
be added in a proper way in order to obtain a gauge invariant result
in the SM.  Unlike in the approach described in the
Section~\ref{tadpoles}, where the new counterterm $Z_v$ for the Higgs VEV
has been introduced, here we include the  tadpole diagrams
explicitely. This makes our renormalisation constants $Z$'s from
Eq. \ref{MSbarren} gauge invariant.

 Below we present the analytical  expressions  for charge and mass \MSb
renormalisation constants,  needed in order to obtain a finite
expression for $\Delta r$ in the \MSb scheme.

\subsection{Coupling and masses renormalisation}

The bare charge $e_0$ and the \MSb charge $e$ are related via 
\begin{eqnarray}
e_0 = \mu^\ep e\:
\left(
1+\frac{e^2(\mu)}{16\pi^2\ep}Z_e^{(1,1)}
+\frac{e^4(\mu)}{(16\pi^2)^2\ep}Z_e^{(2,1)}
+\frac{e^4(\mu)}{(16\pi^2)^2\ep^2}Z_e^{(2,2)}
\right)  \,, \label{msbarcharge}
\end{eqnarray}
where the constants $Z$'s, as we shall see in the following, can depend
on $\sin\theta_W$.

There are two ways to determine the \MSb renormalisation constant in
this  expression. One is to use the Ward--Takahashi identity given in
appendix~\ref{ward} to express it in terms of the gauge boson wave
function renormalisation constants and the renormalisation constant
for $\sin\theta_W$
\begin{eqnarray}
1 = Z_e\left\{ \sqrt{Z_{\gamma\gamma}} +\frac{1}{2}
  \frac{\sin\theta_W^0}{\cos\theta_W^0} \delta Z_{Z\gamma} \right\}.
\end{eqnarray}
Then at the one- and two-loop order for on-shell 
charge renormalisation constants, introduced above, we have
\begin{eqnarray}
\delta Z_{e,\rm OS}^{(1)} &=& -\frac{1}{2} \delta Z_{\gamma\gamma}^{(1)} 
-\frac{1}{2}
\delta Z_{Z\gamma}^{(1)}\frac{\sin\theta_W}{\cos\theta_W},
\\
\delta Z_{e,\rm OS}^{(2)} &=& 
-\frac{1}{2} \delta Z_{\gamma\gamma}^{(2)}
-\frac{1}{2}\frac{\sin\theta_W^0}{\cos\theta_W^0}
\delta Z_{Z\gamma}^{(2)}
+ \left(\delta Z_{e,\rm OS}^{(1)}\right)^2
+\frac{1}{8}\left(\delta Z_{\gamma\gamma}^{(1)}\right)^2.
\end{eqnarray}
Here $\delta Z^{(1,2)}$ are one- and two-loop on-shell field
renormalisation constants, expressed via bare quantities. We can
rewrite $\delta Z^{(2)}_{\gamma\gamma}$ and $\delta Z^{(2)}_{Z\gamma}$ in
terms of self-energy diagrams
\begin{eqnarray}
\delta Z_{\gamma\gamma}^{(2)} &=& 
\Pi_{\gamma\gamma,T}^{(2)'}(0)
+ \delta Z_{\gamma\gamma}^{(1)}\Pi_{\gamma\gamma,T}^{(1)'}(0)
+\delta Z_{Z\gamma}^{(1)}\Pi_{\gamma Z,T}^{(1)'}(0)
-\frac{1}{4}\left(\delta Z_{Z\gamma}^{(1)} \right), \nonumber \\
\delta Z_{Z\gamma}^{(2)} &=&
-\frac{2}{M_Z^2}\left(\Pi_{\gamma Z,T}^{(2)}(0) 
+\frac{1}{2}\delta Z_{Z\gamma}^{(1)}\Pi_{ZZ,T}^{(1)}(0)
+\frac{1}{2}\delta Z_{\gamma\gamma}^{(1)}\Pi_{\gamma Z,T}^{(1)}(0)
+\frac{1}{2}\delta Z_{ZZ}^{(1)}\Pi_{\gamma Z,T}^{(1)}(0)
\right)
-\frac{1}{2}\delta Z_{ZZ}^{(1)}\delta Z_{Z\gamma}^{(1)},
\end{eqnarray}
where this time all of the self energies are unrenormalised.  All
other one-loop field renormalisation constants were defined before in
Section~\ref{ctos}. At the end we have an expression for the on-shell
charge renormalisation constant expressed via bare charge, Weinberg
angle and masses. Now, rewriting the bare quantities in terms of \MSb ones
with yet unknown coefficients in Eq.~\ref{msbarcharge} and requiring
that transition between on-shell and \MSb charge should not contain
divergencies we easily extract the \MSb charge renormalisation constants.

Alternatively, the renormalisation group analysis can be applied.
In order to find $Z_e$ we differentiate Eq. (\ref{msbarcharge})
w.r.t. $\log \mu^2$ and take into account that
\begin{equation}
  \frac{d\,e}{d\,\log\mu^2} = -\frac{\ep}{2}e + \beta_e \,,
\label{chargederivative}
\end{equation}
where 
\begin{equation}
\beta_e = \frac{e^3}{16\pi^2} b_1 + \frac{e^5}{(16\pi^2)^2}b_1+\dots
\end{equation}
is the $\beta$-function. Since $(d/d\log\mu^2)\, e_0=0$,
the l.h.s. of Eq.~\ref{msbarcharge} becomes zero after the
differentiation, while the r.h.s. relates the
coefficients $b_j$ and the unknown constants in (\ref{msbarcharge})
\begin{eqnarray}
Z^{(1,1)}_e  & = & b_{0}, \nonumber\\
Z^{(2,2)}_e  & = & \frac{3}{2} \, b_{0}^2, \nonumber\\
Z^{(2,1)}_e  & = & \frac{1}{2} \, b_{1} \,.
\label{RGZ}
\end{eqnarray}

The function $\beta_e$ can be extracted from the existing calculation
in the unbroken theory. Namely, for the $SU(2)$ and $U(1)$
charges $g$ and $g'$ respectively, the $\beta$-functions read
\begin{eqnarray}
\beta_{g'} & = & \frac{1}{12} \frac{g'^3}{16\pi^2}
    + \frac{1}{4} \frac{g'^5}{(16\pi^2)^2}
    + \frac{3}{4} \frac{g'^3 g^2}{(16\pi^2)^2}\;\:, 
\nonumber \\
\beta_g & = & - \frac{43}{12} \frac{g^3}{16\pi^2}
   - \frac{259}{12} \frac{g^5}{(16\pi^2)^2}
   + \frac{1}{4} \frac{g^3 g'^2}{(16\pi^2)^2}\;\:. 
\label{beta}
\end{eqnarray}
The one-loop result is given in \cite{RG_1loop}, while
the two-loop coefficients have been evaluated in \cite{RG_2loop}.

From the relation
\begin{equation}
\frac{1}{e^2} = \frac{1}{g^2} + \frac{1}{g'^2} \;, 
\label{coupling}
\end{equation}
it is easy to deduce that
\begin{equation}
\beta_e = e^3 \left( \frac{\beta_g}{g^3} 
   + \frac{\beta_{g'}}{g'^3} \right) \,.
\label{beta_e}
\end{equation}

Using now Eqs.~\ref{beta}, \ref{coupling} and \ref{beta_e} we obtain
\begin{equation}
\beta_e = -\frac{7}{2}\frac{e^3}{16\pi^2}
  + \frac{e^5}{(16\pi^2)^2} \left( 
         - \frac{125}{6\sin^2\theta_W} + \frac{1}{2\cos^2\theta_W} \right) ,
\label{beta_e_explicit}
\end{equation}
and, finally, from Eq.~\ref{RGZ} we have
\begin{eqnarray}
Z_e^{(1,1)} &=& -\frac{7}{2} ,\nonumber \\
Z_e^{(2,2)} &=& \frac{147}{8} ,\nonumber \\
Z_e^{(2,1)} &=& -\frac{125}{12}\frac{1}{{\rm sin}^2\theta_{W}}
+\frac{1}{4}\frac{1}{{\rm cos}^2\theta_{W}} .
\label{mschargeres}
\end{eqnarray}
The explicit calculation confirms the above result.

Similarly to the charge renormalisation we write for the masses of the
$Z$, $W$ and the Higgs bosons
\begin{equation}
  (m^0_V)^2 =  m_V^2(\mu)\:
  \left( 1 + \frac{g^2(\mu)}{16\pi^2\ep} Z_V^{(1,1)} 
    + \frac{g^4(\mu)}{(16\pi^2)^2\ep} Z_V^{(2,1)} +
    \frac{g^4(\mu)}{(16\pi^2)^2\ep^2} Z_V^{(2,2)}
  \right). 
  \label{baremsb}
\end{equation}
For $m_Z$ and $m_W$ the renormalisation constants up to two loop
are required while for the Higgs boson we need only the one loop
expression.
The analysis, similar to that described above for the charge,
has been done in details in \cite{olegmisha}. There the explicit
expressions for $Z_V^{(1,1)}$, $Z_V^{(2,1)}$ and $Z_V^{(2,2)}$
are given. 

\subsection{\MSb results for $\Delta r$}

In Fig. 4 we plot $\Delta r^{(2)\overline{\rm MS}}_{\rm bos}$
as a function of the \MSb Higgs boson mass in different scales.
As input parameters we used the on-shell values given in Table I.

\begin{figure}
\epsfig{file=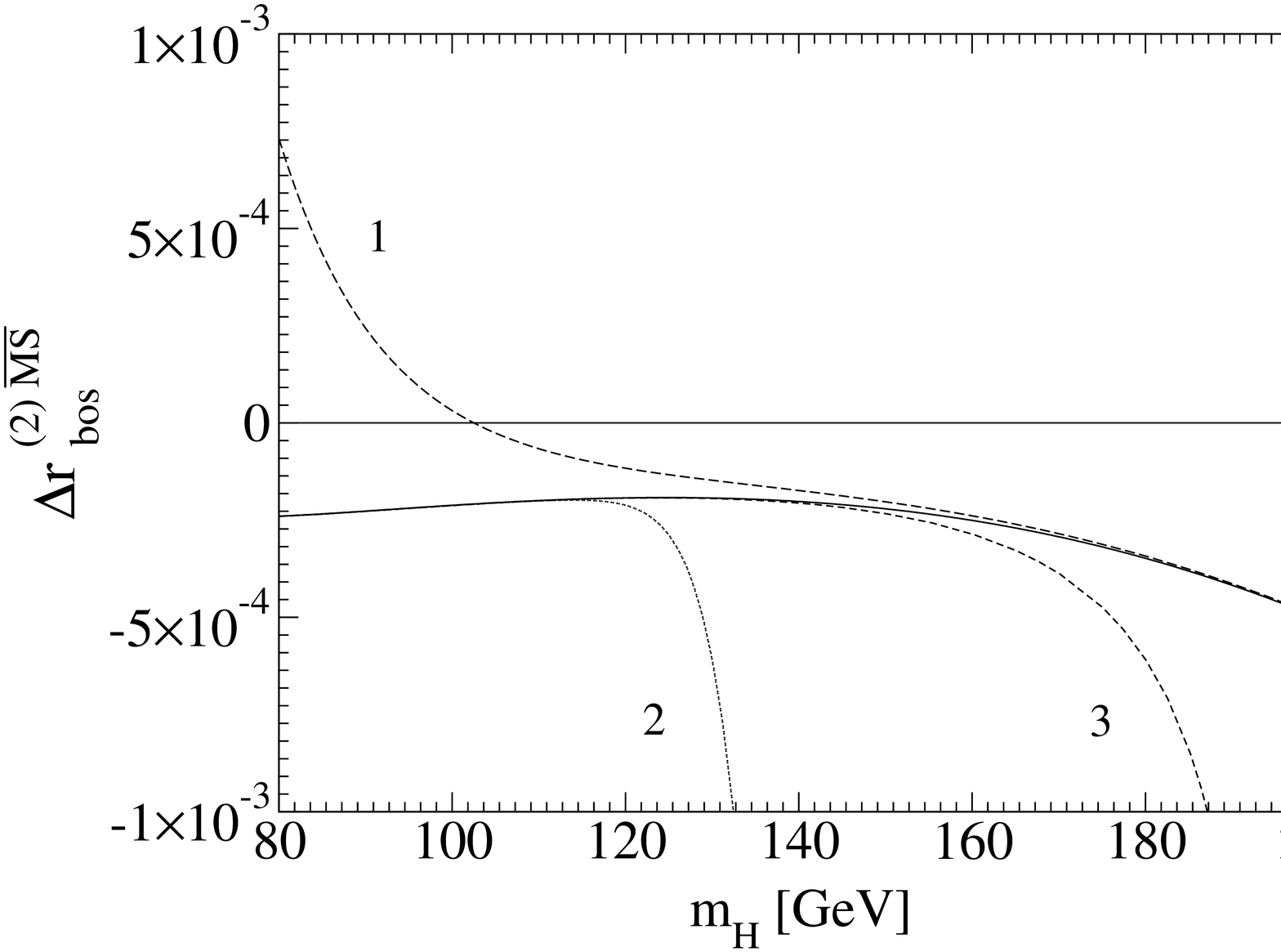,width=6cm,angle=270}
\epsfig{file=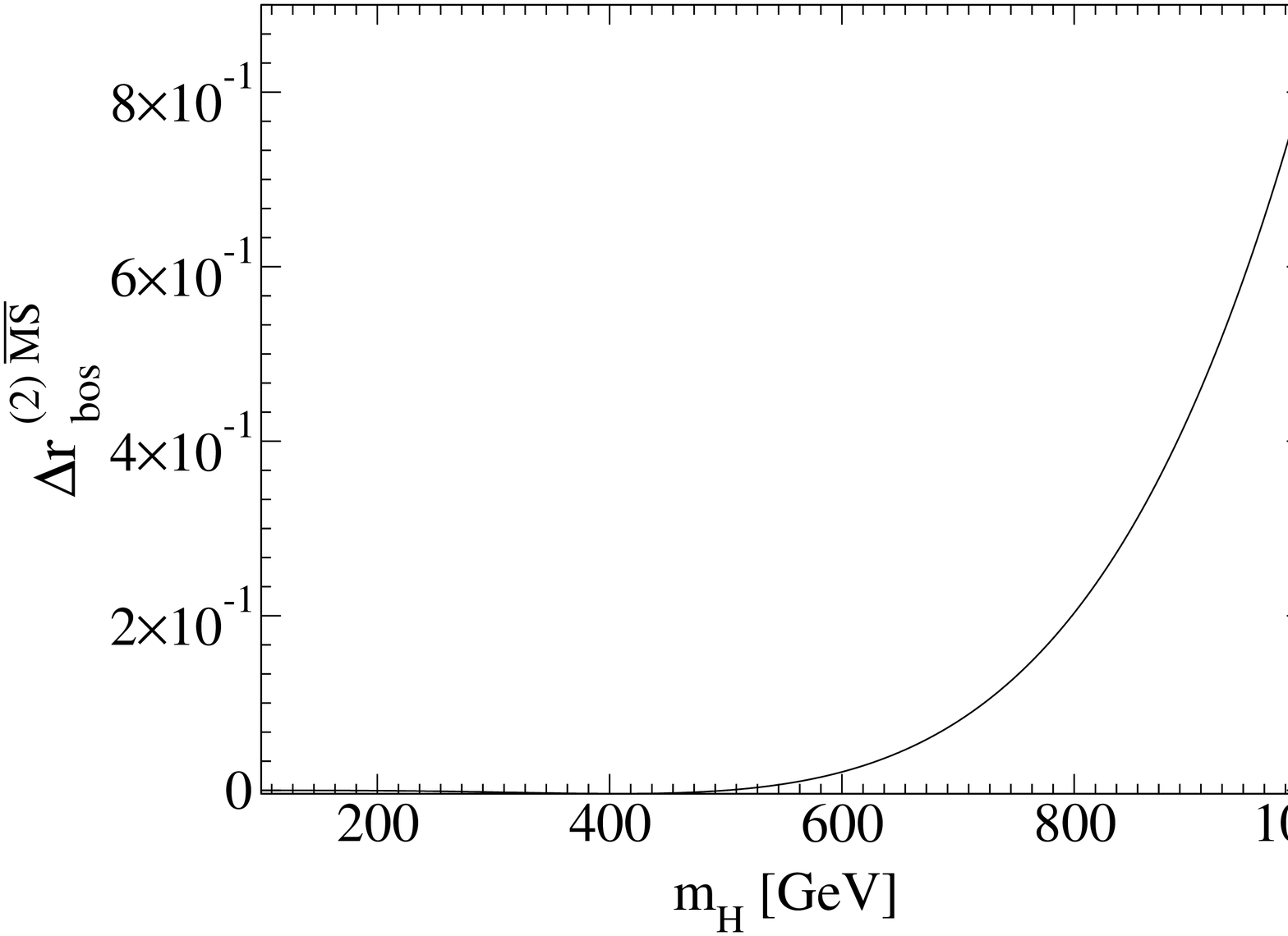,width=6cm,angle=270}
\caption{The exact result (solid line) for the $\Delta r^{(2)}_{\rm bos}$
  in the in the \MSb scheme (left and right panel) vs. its 1) large
  Higgs boson mass expansion (long dashed line) and 2) mass difference
  expansion (dotted line). The short dashed line 3) represents the
  [3/3] Pad\'e approximant.  In the right pannel the large Higgs boson
  mass expansion curve coincides completely with the numerical result
  for this range.}
\end{figure}

The solid curve represents the exact result.
Two other curves represent expansions in different regimes:
as $m_H\to m_Z$ and as $m_H\to\infty$. They cover almost the whole
region of the $m_H$ under consideration. In order to extend the range
of the expansion around $m_Z$ the Pad\'e approximant has been constructed.
It sufficiently improves the situation for the intermediate
Higgs boson masses. Thus the expansions cover completely the region
of interest. The details of expansions are discussed more precisely
in Section~\ref{expansionsdet}.


\section{Transition between the schemes}

\label{transition}

Once we have the result in the \MSb scheme it is necessary to
translate it into the on-shell parameters, which are known with high
precision for the electroweak sector contrary to the strong
interacting sector of the Standard Model. To this end one has to
consider the proper scheme independent quantity which is
\begin{equation}
  \label{invariant}
  \frac{\alpha\pi}{2 M_W^2 \sin^2 \theta_W} (1+\Delta r)
  \equiv \frac{G_F}{\sqrt2} .
\end{equation}
This should be contrasted with the naive approach of taking simply
$\Delta r$ and substituting \MSb parameters.

Using the methods described in Section~\ref{msrenor}, we obtain the
following series expansions connecting on-shell and \MSb parameters
\begin{eqnarray}
  \label{first}
  \alpha_{\OS} &=& \alpha_{\msb} \left(1 + \frac{\alpha_{\msb}}{4 \pi}
  x^\alpha_{1,\msb} + \left( \frac{\alpha_{\msb}}{4 \pi} \right)^2
  x^\alpha_{2,\msb} \right), \\
  M_{W,\OS}^2 &=& M_{W,\msb}^2 \left(1 + \frac{\alpha_{\msb}}{4 \pi}
  x^W_{1,\msb} + \left( \frac{\alpha_{\msb}}{4 \pi} \right)^2
  x^W_{2,\msb} \right), \\
  M_{Z,\OS}^2 &=& M_{Z,\msb}^2 \left(1 + \frac{\alpha_{\msb}}{4 \pi}
  x^Z_{1,\msb} + \left( \frac{\alpha_{\msb}}{4 \pi} \right)^2
  x^Z_{2,\msb} \right), \\
  M_{H,\OS}^2 &=& M_{H,\msb}^2 \left(1 + \frac{\alpha_{\msb}}{4 \pi}
  x^H_{1,\msb} \right). \label{last}
\end{eqnarray}
The series for the Higgs boson mass relation is only needed to first
order, since the Higgs field starts to contribute to the decay only at
the one-loop level. 

\begin{figure}
\epsfig{file=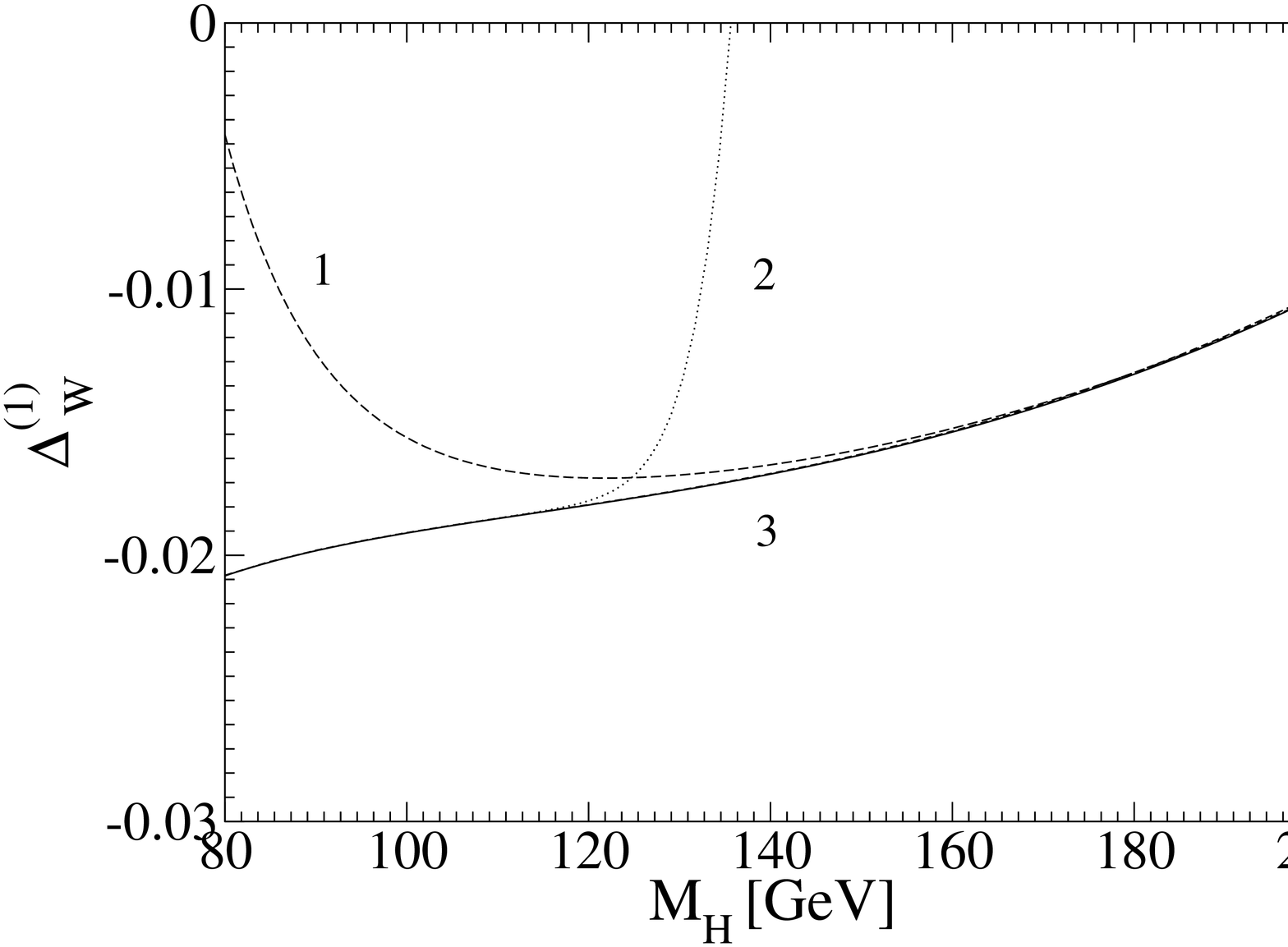,width=6cm,angle=270}
\epsfig{file=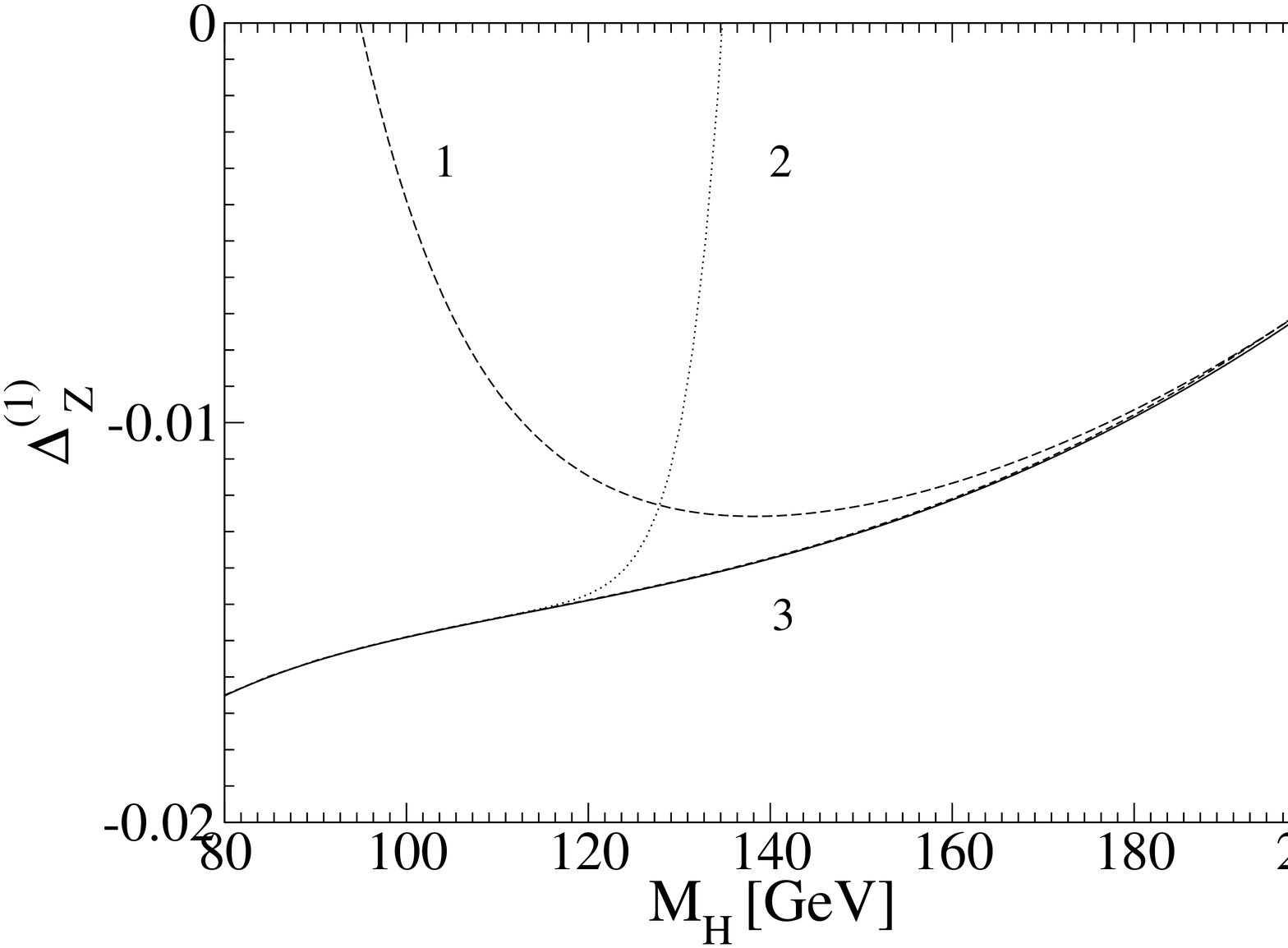,width=6cm,angle=270} \\
\epsfig{file=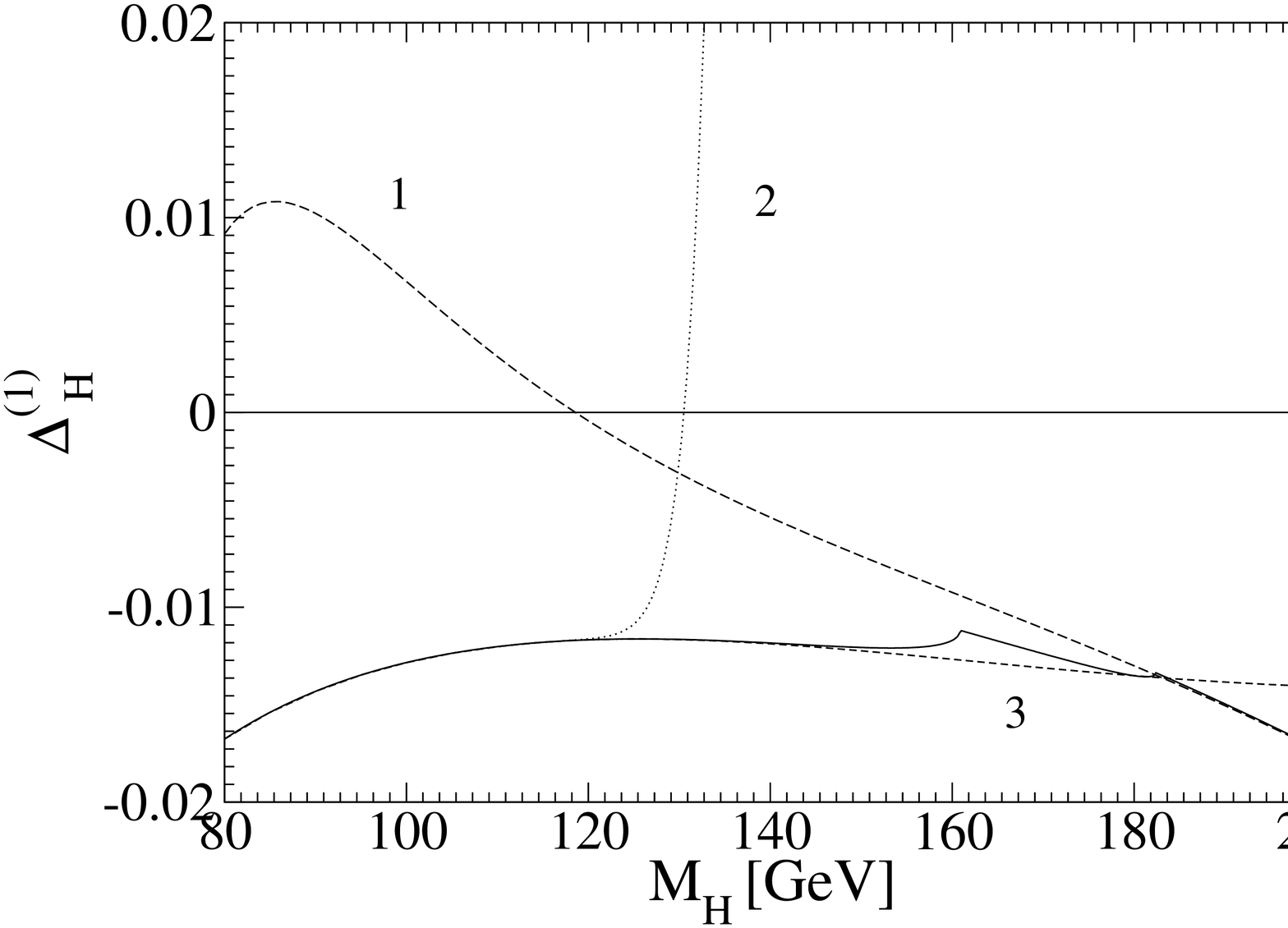,width=6cm,angle=270}
\caption{\label{1loopcoeffs}
  One-loop corrections to the relations between the on-shell and the
  \MSb masses for the $W$, $Z$ and Higgs bosons ($\Delta^1_i =
  \alpha/(4 \pi) X^i_1$). The long dashed line 1) represents the large
  Higgs boson mass expansion, the dotted line 2) represents the mass
  difference expansion. The short dashed line 3) gives the [4/4]
  Pad\'e approximant which coincides for this range with the exact
  result for the $W$ and $Z$ boson mass corrections.}
\end{figure}

The above relations have to be inverted to yield \MSb parameters in
terms of the on-shell ones. For any parameter $A$ the relation will be
written as follows
\begin{equation}
  \label{inverse}
  A_{\msb} = A_{\OS} \left(1 + \frac{\alpha_{\OS}}{4 \pi}
  X^A_{1,\OS} + \left( \frac{\alpha_{\OS}}{4 \pi} \right)^2
  X^A_{2,\OS} \right).
\end{equation}
The expansion coefficients are obtained by inverting the original
series up to the required order. At one-loop this leads trivially to
\begin{equation}
  X^A_{1,\OS} = \left[ -x^A_{1,\msb} \;\; \right]_{M_{i,\msb} \;
  \rightarrow \; M_{i,\OS}}.
\end{equation}
The coefficients for the three bosons are depicted in
Fig.~\ref{1loopcoeffs} with parameters values as given in
Table~\ref{values}, in a comparison of the different evaluation
methods. For Higgs boson masses greater than 200~GeV the large mass
expansion with six coefficients is indiscernible from the numeric
result. The mass difference expansion fails always around 120~GeV.  In
the visible range from 80~GeV to 200~GeV, the Pad\'e approximation
based on the mass difference expansion turns out to practically
coincide with the exact result for the vector bosons. For the Higgs
boson this cannot happen due to the occurrence of the two-particle
production thresholds and indeed there is a region between the
thresholds which cannot be reproduced with neither the mass difference
nor the large mass expansion. Obviously if it was needed this region
could be covered by threshold expansions.

\begin{table}
  \begin{tabular}{|l|l|}
    \hline
    $\alpha^{-1}$ & 137.03599976(50) \\
    $M_W$ & 80.423(39) GeV \\
    $M_Z$ &91.1876(21) GeV \\
    \hline
  \end{tabular} 
  \caption{\label{values}
    Parameter values used in the calculation \cite{Hagiwara:pw}.}
\end{table}

\begin{figure}
\epsfig{file=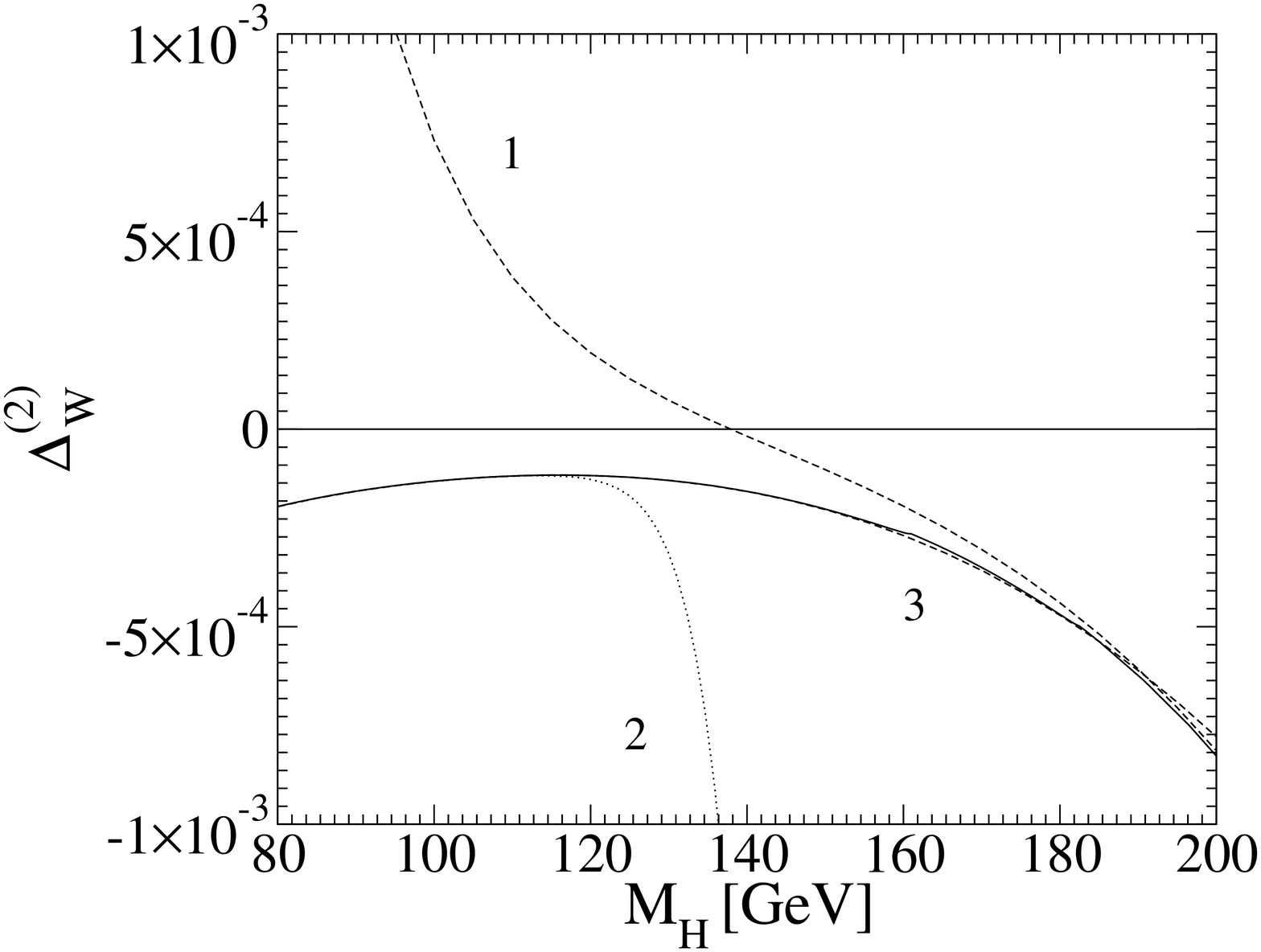,width=6cm,angle=270}
\epsfig{file=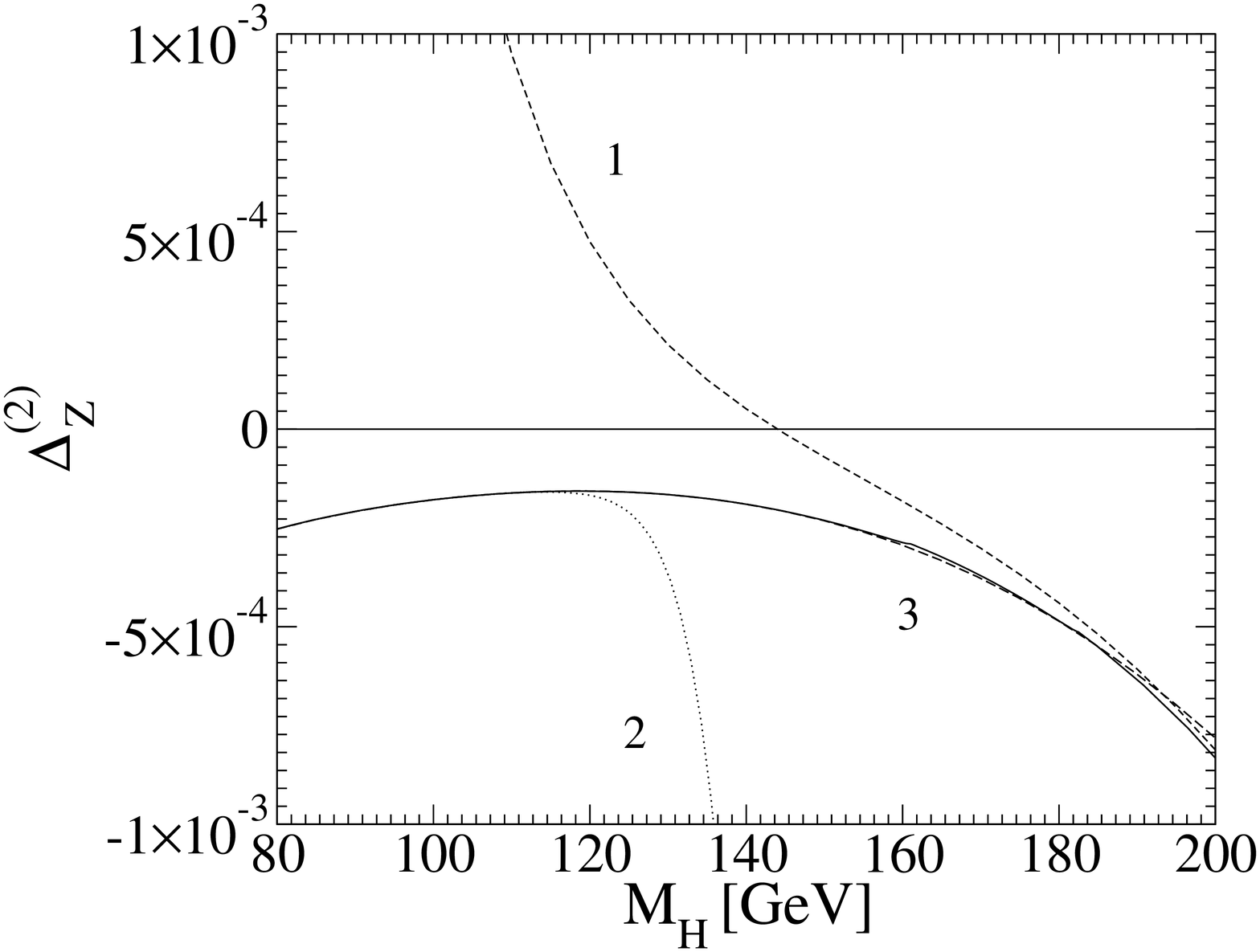,width=6cm,angle=270}
\caption{\label{2loopcoeffs}
  Two-loop corrections to the relation between the on-shell and the
  \MSb masses for the $W$ and $Z$ boson ($\Delta^2_i = (\alpha/(4
  \pi))^2 X^i_2$). The long dashed line 1) represents the large Higgs
  boson mass expansion, the dotted line 2) represents the mass difference
  expansion. The short dashed line 3) gives the [4/4] Pad\'e
  approximant.}
\end{figure}

The two-loop correction contains terms coming also from the one-loop
terms and the proper expression reads
\begin{equation}
  X^A_{2,\OS} = \left[ -x^A_{2,\msb} + x^\alpha_{1,\msb} x^A_{1,\msb}
    + \sum_i M^2_{i,\msb} \frac{\partial x^A_{1,\msb}}{\partial
      M^2_{i,\msb}} x^i_{1,\msb} \;\; \right]_{M_{i,\msb} \;
    \rightarrow \; M_{i,\OS}}.
\end{equation}
The corrections for the vector bosons are depicted similarly to the
one-loop case in Fig.~\ref{2loopcoeffs}. The expansions themselves are
less precise. It is however interesting to note that the Pad\'e
approximation together with the large mass expansion cover the whole
range with high precision. Even the threshold region is reproduced
with a relatively small error, although this is due to the fact that
the peaks are not very pronounced.

\begin{figure}
\epsfig{file=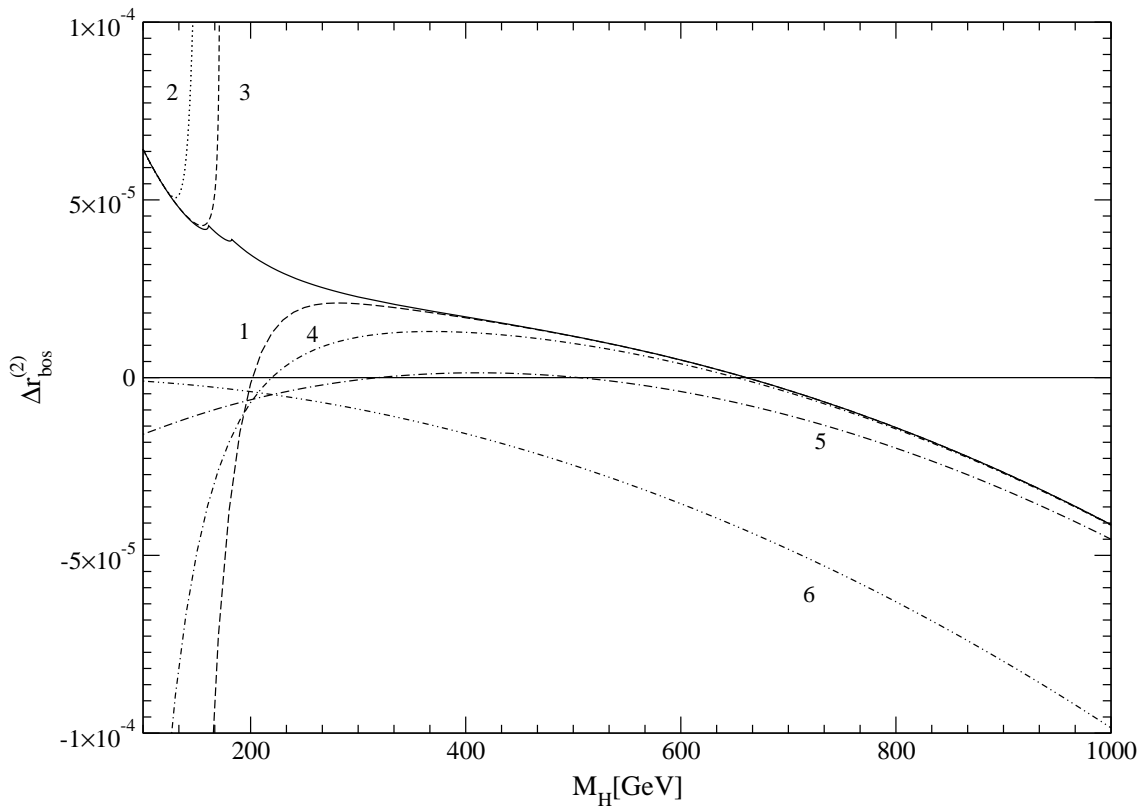,width=11cm,angle=270}
\caption{\label{result}
  On-shell $\Delta r^{(2)}_{\rm bos}$. The long dashed line 1) represents
  the large Higgs boson mass expansion, the dotted line 2) represents
  mass difference expansion. The short dashed line 3) gives the [3/3] Pad\'e
  approximant. The dash-dotted 4) and 5) lines correspond to lower
  terms in the large Higgs boson mass expansion, whereas 6) is the
  leading term.}
\end{figure}

We can now combine all the perturbative expansions and translate the
\MSb result into the on-shell one. We shall not reproduce the formula
since it can be easily obtained from the previous equations. It is
important however to note two things. First, in the expression for the
two-loop $\Delta r$ there are the following terms
\begin{equation}
  \left( \Delta r^{(2)} \right)^{\OS} = \dots
  + \frac{X^W_{2,\OS}}{\sin^2\theta_W} -
  \frac{X^Z_{2,\OS}}{\sin^2\theta_W} + \dots
\end{equation}
If this is combined with the fact that the results in both \MSb and
on-shell schemes behave as $1/\sin^4\theta_W$, it is obvious that one
term in the $W$ and $Z$ boson mass difference expansion is
lost. Second, the result in the \MSb scheme behaves as $M_H^4$,
whereas the one in the on-shell scheme as $M_H^2$. Therefore, one term
in large Higgs boson mass expansion is also lost. As a result, if the
expansions of \cite{olegmisha} are taken, the final result can be
given with five coefficients in both expansions in the large mass
case. The formulae can be found in Appendix~\ref{onshell}. The mass
difference expansion requires an independent calculation of the
on-shell propagator diagrams and the result can be found in
Appendix~\ref{onshelldiff}. The numeric results can be found in
Fig.~\ref{result}. It should be stressed, that it was checked that the
exact analytic result without expansions obtained by the translation
procedure described above and by an explicit renormalisation in the
on-shell scheme are the same.

\begin{figure}
\epsfig{file=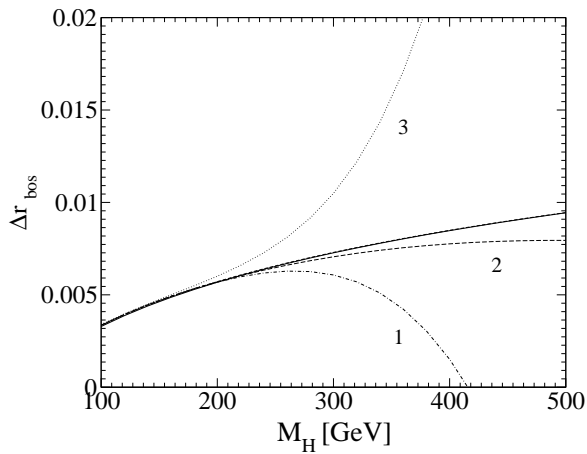,width=6cm,angle=270}
\caption{\label{naive}
Numeric translation of $\Delta r$ form the \MSb scheme to the on-shell
scheme vs. the exact result (solid line).}
\end{figure}

It is interesting to consider the transition between the schemes
performed purely numerically. In Fig.~\ref{naive}, the solid curve
represents the one-loop correction as well as the sum of the one- and
two-loop corrections. The fact that they are indiscernible in this
scale is due to their relative smallness. The most reliable way of
obtaining the correction (apart from the exact method) is to take the
one-loop result and substitute the \MSb parameters only in the
normalisation in Eq.~\ref{invariant}, whereas the masses in 
$(\Delta r^{(1)}_{\text{bos}})^{\msb}$ should be left in the on-shell
scheme. This is shown in the curve 2). If one, however, simply takes
the whole invariant and substitutes all of the \MSb parameters, then
curve 1) is obtained, which diverges strongly for Higgs boson masses
larger than about 250~GeV. It turns out that the sum
of the one- and two-loop corrections does not reduce substantially the
scheme dependence, as shown by curve 3), where the correction up to
two-loop order in the \MSb scheme has been given for \MSb parameters
translated from on-shell values using Eqs.~\ref{first} to \ref{last}.

\section{Computational methods}

The calculation of the bosonic corrections to the muon life time is a
relatively complex task. The number of Feynman diagrams to be
calculated is around 5000 in $R_\xi$ gauge.  This makes it necessary
to use automated software.

\subsection{Software and checks}

The first step of the calculation is the generation of
diagrams. Several systems are presently available. Obviously each
differs in its easy of use, speed and design concepts. 

  The on-shell calculation was based on the C++ library 
{\bf DiaGen} \cite{diagen}.
It generates all diagrams together with all necessary counterterms.
The main advantage of this software is the speed, since all of the
diagrams were generated in a few seconds, thus making the generation
phase a negligible part of the calculation.

Alternatively, for the calculation with the tadpoles the input
generator {\bf DIANA} \cite{DIANA} has been applied. We note that according
to the rules, given in Section~\ref{msrenor}, no counterterm diagrams
should be generated. They are all taken into account by the
multiplicative renormalisation.

The diagrams to be evaluated can be divided into two broad
classes. First are these which can be reduced to vacuum bubbles. Here,
partial integration identities \cite{Chetyrkin:qh} supplied with
analytical formulae \cite{Davydychev:1992mt} can be used.

The second more complicated problem is the evaluation of the two-loop
two-point functions at non-vanishing external momentum (at the values
$q^2=M_Z^2$ and $q^2=M_W^2$ in our case).  From the several
possibilities two different algorithms have been used to deal with
these diagrams.

The algorithm described in \cite{Weiglein:hd} has been chosen because
of its simplicity. As an end result of the tensor reduction scalar
two-loop propagator integrals are obtained. A high precision numerical
evaluation of these is currently possible with one dimensional
integral representations \cite{Bauberger:1994hx}. To this end C++
programs were used based on the library {\bf S2LSE} \cite{s2lse}. For
large scale differences which occur when the Higgs mass is much above
the masses of the $W$ and the $Z$ boson double precision turns out to
be insufficient. An easy way to see it is to remark that the
individual terms in the result can behave as $M_H^8$ whereas due to
the screening theorem \cite{Veltman:1976rt} the whole result behaves
at most as $M_H^2$. For a Higgs boson mass of the order of 1~TeV, this
means that cancellations of the order of $10^6$ will have to occur. If
we combine this with the fact that in double precision some of the
integrals can only be evaluated to 5 digits, the numerical instability
becomes apparent. A way out of this problem on 32 bit machines is to
use software emulated quadruple precision. Of course this signifies an
important drop in effectiveness. In practice, the software runs about
20 times slower. Ten times are due to the use of software emulation
for arithmetical operations and two to more integration points which
are needed for higher precision. On present GHz processors, the
evaluation of a single point of the final result requires around 20s
and a conservative estimate of the error over the whole range of Higgs
boson mass from 100~GeV to 1~TeV is four digits.

Alternatively to the numerical method, we used also the semianalytic
method of expansions (see next subsection). In this case the huge
cancellations mentioned above do not cause any problem.

\begin{figure}
\epsfig{file=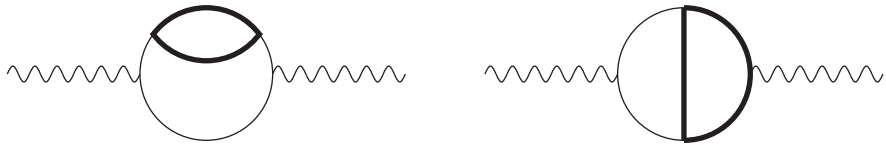}
\caption{\label{LargeMass}
A choice of formally heavy lines in the large mass expansion of two
$W$ boson propagator topologies.}
\end{figure}

The size of the programs written in C++ and in {\bf FORM}
\cite{Vermaseren:2000nd} requires stringent checks. A helpful property
of the bosonic corrections to the propagators is that the value of
every single diagram can be obtained rather easily through low
momentum or large mass expansions. In fact for the $Z$ boson
propagators a low momentum expansion up to tenth order provided a five
digit agreement with the integral representations for each diagram
independently and for the whole sum. Additionally, we also made  an
expansion around the point $M_H=M_Z$ (see next subsection) and got
excellent agreement between the numerical and the expanded results.
In the case of the $W$ boson propagators not all of the diagrams are
below threshold. It turns out that 345 contain a photon or a massless
ghost line, which makes as much as around 160 of them to be either on
threshold or infrared divergent. In this case the low momentum
expansion either fails to converge or converges very slowly. A way out
of this is given by large mass expansions. If the lines which are to
be considered as heavy are chosen in a specific way, then the large
mass expansion leads only to vacuum bubbles and one-loop propagator
diagrams and the convergence is comparable to the case of the $Z$
boson propagators. An example choice of the heavy lines for two
different topologies is given in Fig.~\ref{LargeMass}. This procedure
fails only for graphs which represent pure $QED$ corrections to a $W$
boson line. In this case however, the result is known analytically
\cite{Gray:1990yh}.

Another way of testing the analytical reduction and the diagram
generation software is to check the Ward--Takahashi identities for the
propagators. Here the following relations have been evaluated
\begin{eqnarray}
  p^2 \left( \Pi^{(2)}_{ZZ,L} + 2 i M_Z \Pi^{(2)}_{Z G_Z} \right)
   + M_Z^2 \Pi^{(2)}_{G_Z G_Z} - p^2 \left( \Pi^{(1)}_{Z G_Z} \right)^2
   + \Pi^{(1)}_{ZZ,L}  \Pi^{(1)}_{G_Z G_Z} &=& 0, \\
  p^2 \left( \Pi^{(2)}_{WW,L} - 2 M_W \Pi^{(2)}_{W G_W} \right)
   + M_W^2 \Pi^{(2)}_{G_W G_W} - p^2 \left( \Pi^{(1)}_{W G_W} \right)^2
   + \Pi^{(1)}_{WW,L}  \Pi^{(1)}_{G_W G_W} &=& 0
\end{eqnarray}
both for on-shell values of the momentum and in an expansion around
zero up to third order. 
Here $G_Z$ and $G_W$ stand for the neutral and charged would-be
Goldstone boson respectively and
the subscript ``$L$'' denotes longitudinal parts
of the vector boson self-energies and the scalar vector transitions
are given by
\begin{equation}
  \Pi_{V G_V}^\mu (p) = p^\mu \Pi_{V G_V} (p^2),
\end{equation}
where $p$ is the ingoing momentum of the vector boson.

The combination of the two checks described
above, tests the software from the diagram generation to the numerical
evaluation. An additional test is of course provided by gauge
invariance and indeed the calculation was performed in the general
$R_\xi$ gauge with three independent gauge parameters. We have observed
explicitely the cancellation of each of them from the final result and
the counterterms.

Since the bosonic corrections to the propagators in the
\MSb scheme have been evaluated within the large
Higgs boson mass approach in \cite{olegmisha} a comparison
was also possible for the whole result. It turns out that the
agreement is perfect for Higgs masses running as low as 200~GeV.

To complete the description of the computational methods, let us note
that C++ and {\bf FORM} were supplied with a collection of AWK and
Bourne shell scripts managed by several Makefiles. The system prepared
in this way runs completely automatically from the beginning with
diagram generation up to the numerical evaluation with
plots. Actually, the specificity of the problem allowed to reduce the
evaluation time of the whole problem down to only one hour and a half,
which is rather short for multiloop calculations.

\subsection{Expansions}

\label{expansionsdet}
  
Here we give more details on how  the expansions are performed 
in two different regimes that we considered
\begin{itemize}
\item in the mass difference $h_Z=(M_H^2-M_Z^2)/M_Z^2$ and
\item in the mass ratio $z_H=M_Z^2/M_H^2$.
\end{itemize}

The expansion in the mass difference $M_H^2-M_Z^2$ is especially
simple. It is just a Taylor expansion of all Higgs propagators and Higgs
boson masses in the vertices around $M_Z$.  No additional subgraphs are
necessary in this case.  The expansion in the heavy Higgs boson limit
is somewhat more involved. It is given by the rules of asymptotic
expansions \cite{asymptotic}.

In addition in the presence of both $M_Z$ and $M_W$ we expand in the
difference of these masses as well. Indeed
\begin{equation}
  \frac{M_Z^2-M_W^2}{M_Z^2} = \sin^2\theta_W \approx 0.23 ,
\end{equation}
is a rather small parameter and the convergence of this series is
quite fast. This trick was used previously in \cite{olegmisha}.  The
advantage of this approach is that in the case of on-shell Green
functions all integrals have only one scale. This allows one to use
the {\bf FORM} package {\bf ONSHELL2} \cite{ONSHELL2} 
to evaluate these integrals analytically.

We should also note that to extend the range of the $h_Z$ expansion we
apply the Pad\'e approximation. Throughout this paper we use a [3/3]
Pad\'e approximant for $\Delta r$ and [4/4] for the scheme transition
formulae. The Pad\'e approximation for the $z_H$ series does not work
well since this series is nonalternating.


\section{Conclusions}

The recent calculation of the two-loop bosonic corrections to $\Delta
r$ performed by two independent groups has been described in detail,
from the matching onto the Fermi theory to the renormalisation and the
explicit results in the on-shell and \MSb schemes. 
The framework for the evaluation of the Fermi constant $G_F$ based
on the low energy factorisation theorem has been constructed.
It allows one to compute $G_F$ as a Wilson coefficient in a simple manner. 
This approach is general and is also applicable to other low energy quantities.

A comparison of different expansions and numerical methods  has been
given. It has been proven that in the wide range of Higgs  boson
masses expansions provide as much precision as needed and cover  the
whole region of interest. The only problematic region, however,  is
connected to the thresholds for $W$ and $Z$ boson pair production. If
the Higgs boson was indeed found in this range, then a precise result
could also be obtained with expansions but this time of the threshold
type. The coincidence of the numerical and analytical results serves
as a strong check of the calculation.

The accuracy of the numerical transformation between \MSb and on-schemes
has been tested. It is shown that for the Higgs boson masses larger
than $\sim250$ GeV the two loop correction does not reduce the scheme
dependence which can be explained by huge cancelations of large terms
during the transition procedure.


\section{Acknowledgments}

The authors would like to thank K. Chetyrkin for fruitful
discussions. A. O and O. V. thank M. Tentyukov for his help with DIANA.
M. A. would like to thank the ``Marie Curie Programme''
of the European Commission for a stipend.  M. C. would like to thank
the Alexander von Humboldt foundation for fellowship. This work was
supported in part by the European Community's Human Potential
Programme under contract HPRN-CT-2000-00149 Physics at Colliders, 
by the KBN Grant 5P03B09320,
by DFG-Forschergruppe  {\it ``Quantenfeldtheorie,
               Computeralgebra und\\ Monte-Carlo-Simulation''}
               (contract FOR 264/2-1) and
by BMBF under grant No 05HT9VKB0.


\appendix


\section{$U(1)$ Ward identity and the renormalisation of charge}

\label{ward}

In this appendix we present a derivation of the relation between the
charge renormalization constant and different wave function
renormalization constants valid to all orders of perturbation theory.
The derivation is based on the use of the $U(1)$ Ward--Takahashi
identity for the weak hypercharge gauge group. To begin with, let us take
the bare $U(1)$ gauge boson field $B_{\mu}^0$ and rewrite it in terms
of mass eigenstates
\begin{eqnarray}
  B_{\mu}^0 = c_W^0A_{\mu}^0 + s_W^0Z_{\mu}^0.
\end{eqnarray}
Here $c_W^0 = \cos\theta_W^0$ and  $s_W^0 = \sin\theta_W^0$ are bare
values  of cosine and sine of Weinberg angle.

In the next step we express our bare gauge boson fields through
the renormalized ones
\begin{eqnarray}
  \left( Z_2^B\right)^{1/2}\left\{c_W A_{\mu} + s_W Z_{\mu}\right\}
  = c_W^0 \left\{\frac{1}{2} Z_{\gamma Z} Z_{\mu} 
    +\left(Z_{\gamma\gamma}\right)^{1/2}A_{\mu}\right\}
  +s_W^0\left\{\left(Z_{ZZ}\right)^{1/2} Z_{\mu} 
    + \frac{1}{2} Z_{Z\gamma }A_{\mu} \right\}.
\end{eqnarray}
Now taking the coefficient in front of $A_{\mu}$ in the equation above
we have
\begin{eqnarray} 
  \left(Z_2^B\right)^{1/2} c_W = c_W^0\left(Z_{\gamma\gamma}\right)^{1/2}
  +\frac{1}{2} s_W^0 Z_{Z\gamma }. \label{Wardinterm}
\end{eqnarray}
To complete the derivation we need to relate the $Z_2^B$ renormalisation
constant to the charge renormalization constant. The electric charge is
related to the weak hypercharge via the following equation
\begin{eqnarray}
  e^0 = g_1^0 c_W^0 = \left(Z_2^B\right)^{-1/2}g_1 c_W^0 =
  Z_e e = Z_e g_1 c_W,
\end{eqnarray}
where we have made use of $U(1)$ Ward--Takahashi identity 
$Z_{g_1} = \left(Z_2^B\right)^{-1/2}$. Now we can easily deduce, that
\begin{eqnarray}
Z_e = \left(Z_2^B\right)^{-1/2}\frac{c_W^0}{c_W}.
\end{eqnarray}
Substituting this relation into Eq.~\ref{Wardinterm} we have
\begin{eqnarray}
1 = Z_e\left\{\left(Z_{\gamma\gamma}\right)^{1/2}
+\frac{s_W^0}{c_W^0}\frac{1}{2} Z_{Z\gamma }\right\}.
\end{eqnarray}
Using this final relation one can considerably simplify the
calculation of the on-shell charge renormalization constant and  avoid
dealing with infrared rearrangement while computing the three-point
Green function.


\section{Large Higgs boson mass expansion of $\Delta
  r^{(2)}_{\text{bos}}$ in the on-shell scheme}

\label{onshell}

In this appendix, the on-shell renormalised $\Delta
r^{(2)}_{\text{bos}}$ is given in a twofold expansion, in the large
Higgs boson mass and in the mass difference between the $W$ and the
$Z$ boson. The 
number of terms is consistent with the result
\cite{olegmisha} as explained in
Section~\ref{transition}. The leading behaviour both in the Higgs mass
and in the sine of the Weinberg angle has been factorised out.
\begin{equation}
  \left( \Delta r^{(2)}_{\text{bos}} \right)^{\OS} = \left( 
    \frac{\alpha}{4 \pi \sin^2 \theta_W}\right)^2 \frac{M_H^2}{M_Z^2}
    \sum^{4}_{n=0} \sin^{2n} \theta_W R^{\OS}_n.
\end{equation}
The occurring transcendental numbers are
\begin{eqnarray}
  \label{transcendental}
  S_1 &=& \frac{\pi}{\sqrt{3}}, \\
  S_2 &=& \frac{4}{9}
  \frac{\text{Cl}_2(\pi/3)}{\sqrt{3}} \simeq
  0.2604341376321620989557291432080308... \nonumber
\end{eqnarray}
while $z_H = M_Z^2/M_H^2$. Note that the leading term in the Higgs
boson mass can be resumed in $\sin^2 \theta_W$ to give the behaviour
\begin{equation}
  \Delta r^{(2)}_{\text{Higgs}} = \left( \frac{\alpha}{4 \pi \sin^2
  \theta_W} \right)^2 \frac{M_H^2}{8 M_W^2} \left( 9 \sqrt{3} \mbox{
  Cl}_2 (\frac{\pi}{3}) + \frac{49}{72} - \frac{11 \pi \sqrt{3}}{4} -
  \frac{25 \pi^2}{108} \right).
\end{equation}
The expansion coefficients read (the first four of them expanded to the order
$O(z_H^3)$ were already published in \cite{oni})
\begin{eqnarray}
\nonumber
R^{\OS}_0=
 & & \Biggl[
-\tfrac{33}{32} \, S_1
-\tfrac{25}{144} \, \zeta_2
+\tfrac{49}{576}
+\tfrac{243}{32} \, S_2
\Biggr] \nonumber \\
&+&z_H \, \Biggl[
-\tfrac{303659}{384} \, S_1
-\tfrac{1515}{16} \, \zeta_3
-\tfrac{125}{4} \, \text{ln}(z_H) \, S_1
+\tfrac{41}{288} \, \text{ln}(z_H)^2
+\tfrac{17305}{576} \, \zeta_2
+\tfrac{98125}{1728} \, \text{ln}(z_H)
+\tfrac{4131}{32} \, S_1 \, S_2
\nonumber \\ & & 
+\tfrac{1089}{8} \, S_1^2
+\tfrac{7069829}{20736}
+\tfrac{259443}{128} \, S_2
+66 \, \text{ln}(3) \, \zeta_2
\Biggr] \nonumber \\ & & 
\nonumber \\ &+&z_H^2 \, \Biggl[
-\tfrac{75341}{960} \, S_1
-\tfrac{41441}{1440} \, \zeta_2
-\tfrac{83}{48} \, \text{ln}(z_H)^2
+\tfrac{5940941}{86400}
+\tfrac{42481}{320} \, \text{ln}(z_H)
+\tfrac{72171}{320} \, S_2
-95 \, \text{ln}(z_H) \, S_1
\Biggr] \nonumber \\ & & 
\nonumber \\ &+&z_H^3 \, \Biggl[
-\tfrac{1519543}{3200} \, S_1
-\tfrac{29439}{80} \, \text{ln}(z_H) \, S_1
-\tfrac{4031}{30} \, \zeta_2
-\tfrac{70567}{2880} \, \text{ln}(z_H)^2
+\tfrac{6319637}{17280} \, \text{ln}(z_H)
+\tfrac{599726311}{1296000}
+\tfrac{95067}{160} \, S_2
\Biggr] \nonumber \\ & & 
\nonumber \\ &+&z_H^4 \, \Biggl[
-\tfrac{34374449}{16800} \, S_1
-\tfrac{7468}{5} \, \text{ln}(z_H) \, S_1
-\tfrac{1837867}{3360} \, \zeta_2
-\tfrac{92423}{720} \, \text{ln}(z_H)^2
+\tfrac{11836033}{10800} \, \text{ln}(z_H)
+\tfrac{3853287}{2240} \, S_2
+\tfrac{47950477091}{25401600}
\Biggr] \nonumber \\
R^{\OS}_1=
 & & \Biggl[ 
-\tfrac{33}{32} \, S_1
-\tfrac{25}{144} \, \zeta_2
+\tfrac{49}{576}
+\tfrac{243}{32} \, S_2
\Biggr] \nonumber \\
&+&z_H \, \Biggl[
-\tfrac{772245}{128} \, S_2
-\tfrac{4850423}{2304}
-\tfrac{8349}{16} \, S_1^2
-\tfrac{14733}{32} \, S_1 \, S_2
-\tfrac{44117}{192} \, \zeta_2
-\tfrac{24641}{192} \, \text{ln}(z_H)
+\tfrac{115}{288} \, \text{ln}(z_H)^2
\nonumber \\ & & 
+\tfrac{6541}{96} \, \text{ln}(z_H) \, S_1
+\tfrac{5653}{16} \, \zeta_3
+\tfrac{3844355}{1152} \, S_1
-264 \, \text{ln}(3) \, \zeta_2
-66 \, \text{ln}(3) \, S_1
+96 \, \text{ln}(2) \, \zeta_2
\Biggr] \nonumber \\ & & 
\nonumber \\ &+&z_H^2 \, \Biggl[
-\tfrac{373487}{960} \, \text{ln}(z_H)
-\tfrac{7519453}{28800}
-\tfrac{40167}{160} \, S_2
+\tfrac{203}{72} \, \text{ln}(z_H)^2
+\tfrac{47807}{720} \, \zeta_2
+\tfrac{248651}{1440} \, S_1
+\tfrac{11653}{48} \, \text{ln}(z_H) \, S_1
\Biggr] \nonumber \\ & & 
\nonumber \\ &+&z_H^3 \, \Biggl[
-\tfrac{5103328117}{2592000}
-\tfrac{108967649}{86400} \, \text{ln}(z_H)
-\tfrac{22023}{80} \, S_2
+\tfrac{189529}{2880} \, \text{ln}(z_H)^2
+\tfrac{3561}{10} \, \zeta_2
+\tfrac{168033}{160} \, \text{ln}(z_H) \, S_1
+\tfrac{13792549}{9600} \, S_1
\Biggr] \nonumber \\ & & 
\nonumber \\ &+&z_H^4 \, \Biggl[
-\tfrac{11560206263}{1296000}
-\tfrac{1265074333}{302400} \, \text{ln}(z_H)
+\tfrac{122819}{288} \, \text{ln}(z_H)^2
+\tfrac{44091}{56} \, S_2
+\tfrac{4466227}{2520} \, \zeta_2
+\tfrac{58457}{12} \, \text{ln}(z_H) \, S_1
+\tfrac{72817723}{10080} \, S_1
\Biggr] \nonumber \\
R^{\OS}_2=
 & & \Biggl[ 
-\tfrac{33}{32} \, S_1
-\tfrac{25}{144} \, \zeta_2
+\tfrac{49}{576}
+\tfrac{243}{32} \, S_2
\Biggr] \nonumber \\
&+&z_H \, \Biggl[
-\tfrac{1857047}{384} \, S_1
-\tfrac{17687}{48} \, \zeta_3
-\tfrac{635}{32} \, \text{ln}(z_H) \, S_1
+\tfrac{115}{288} \, \text{ln}(z_H)^2
+\tfrac{21407}{576} \, \text{ln}(z_H)
+\tfrac{11109}{32} \, S_1 \, S_2
+\tfrac{256511}{576} \, \zeta_2
\nonumber \\ & & 
+\tfrac{4719}{8} \, S_1^2
+\tfrac{535053}{128} \, S_2
+\tfrac{94100843}{20736}
-192 \, \text{ln}(2) \, \zeta_2
+242 \, \text{ln}(3) \, S_1
+288 \, \text{ln}(3) \, \zeta_2
\Biggr] \nonumber \\ & & 
\nonumber \\ &+&z_H^2 \, \Biggl[
-\tfrac{172439}{960} \, S_1
-\tfrac{2539}{16} \, \text{ln}(z_H) \, S_1
-\tfrac{33067}{720} \, \zeta_2
+\tfrac{11}{12} \, \text{ln}(z_H)^2
+\tfrac{32847}{160} \, S_2
+\tfrac{2467333}{8640} \, \text{ln}(z_H)
+\tfrac{165415279}{518400}
\Biggr] \nonumber \\ & & 
\nonumber \\ &+&z_H^3 \, \Biggl[
-\tfrac{1135551}{640} \, S_1
-\tfrac{8525}{8} \, \text{ln}(z_H) \, S_1
-\tfrac{23417}{80} \, \zeta_2
-\tfrac{12427}{576} \, \text{ln}(z_H)^2
+\tfrac{19761}{40} \, S_2
+\tfrac{136065731}{86400} \, \text{ln}(z_H)
+\tfrac{590467441}{216000}
\Biggr] \nonumber \\ & & 
\nonumber \\ &+&z_H^4 \, \Biggl[
-\tfrac{189343717}{16800} \, S_1
-\tfrac{32914}{5} \, \text{ln}(z_H) \, S_1
-\tfrac{5087309}{2520} \, \zeta_2
-\tfrac{116609}{360} \, \text{ln}(z_H)^2
-\tfrac{675}{112} \, S_2
+\tfrac{90743033}{12096} \, \text{ln}(z_H)
+\tfrac{139434492997}{9072000}
\Biggr] \nonumber \\
R^{\OS}_3=
 & & \Biggl[ 
-\tfrac{33}{32} \, S_1
-\tfrac{25}{144} \, \zeta_2
+\tfrac{49}{576}
+\tfrac{243}{32} \, S_2
\Biggr] \nonumber \\
&+&z_H \, \Biggl[
-\tfrac{29707045}{10368}
-\tfrac{77817}{64} \, S_2
-\tfrac{366221}{864} \, \zeta_2
-\tfrac{5071}{24} \, S_1^2
-\tfrac{589}{3} \, \text{ln}(3) \, S_1
-\tfrac{664}{9} \, \text{ln}(3) \, \zeta_2
-\tfrac{709}{16} \, S_1 \, S_2
\nonumber \\ & & 
-\tfrac{5257}{1728} \, \text{ln}(z_H)
+\tfrac{115}{288} \, \text{ln}(z_H)^2
+\tfrac{1739}{864} \, \text{ln}(z_H) \, S_1
+\tfrac{10357}{72} \, \zeta_3
+\tfrac{1769833}{648} \, S_1
\Biggr] \nonumber \\ & & 
\nonumber \\ &+&z_H^2 \, \Biggl[
-\tfrac{46220713}{259200}
-\tfrac{36923}{960} \, \text{ln}(z_H)
+\tfrac{11}{12} \, \text{ln}(z_H)^2
+\tfrac{2221}{108} \, \text{ln}(z_H) \, S_1
+\tfrac{31091}{1440} \, \zeta_2
+\tfrac{1428847}{25920} \, S_1
+\tfrac{41379}{320} \, S_2
\Biggr] \nonumber \\ & & 
\nonumber \\ &+&z_H^3 \, \Biggl[
-\tfrac{2086198469}{1036800}
-\tfrac{35333761}{43200} \, \text{ln}(z_H)
+\tfrac{12311}{2880} \, \text{ln}(z_H)^2
+\tfrac{109387}{960} \, \zeta_2
+\tfrac{121301}{640} \, S_2
+\tfrac{110077}{240} \, \text{ln}(z_H) \, S_1
+\tfrac{45009703}{43200} \, S_1
\Biggr] \nonumber \\ & & 
\nonumber \\ &+&z_H^4 \, \Biggl[
-\tfrac{658300265597}{42336000}
-\tfrac{227753417}{33600} \, \text{ln}(z_H)
+\tfrac{9103}{80} \, \text{ln}(z_H)^2
+\tfrac{623873}{840} \, S_2
+\tfrac{99685}{84} \, \zeta_2
+\tfrac{2489969}{540} \, \text{ln}(z_H) \, S_1
+\tfrac{4336498723}{453600} \, S_1
\Biggr] \nonumber \\
R^{\OS}_4=
 & & \Biggl[ 
-\tfrac{33}{32} \, S_1
-\tfrac{25}{144} \, \zeta_2
+\tfrac{49}{576}
+\tfrac{243}{32} \, S_2
\Biggr] \nonumber \\
&+&z_H \, \Biggl[
-\tfrac{52145731}{77760} \, S_1
-\tfrac{1589}{32} \, S_1 \, S_2
-\tfrac{7069}{144} \, S_1^2
-\tfrac{5563}{144} \, \zeta_3
-\tfrac{8165}{2592} \, \text{ln}(z_H) \, S_1
+\tfrac{115}{288} \, \text{ln}(z_H)^2
+\tfrac{49141}{8640} \, \text{ln}(z_H)
\nonumber \\ & & 
+\tfrac{350}{9} \, \text{ln}(3) \, S_1
+\tfrac{404}{9} \, \text{ln}(3) \, \zeta_2
+\tfrac{689639}{1920} \, S_2
+\tfrac{1039343}{2880} \, \zeta_2
+\tfrac{6966953}{11520}
\Biggr] \nonumber \\ & & 
\nonumber \\ &+&z_H^2 \, \Biggl[
-\tfrac{4768757}{103680}
-\tfrac{16457}{1296} \, \text{ln}(z_H) \, S_1
-\tfrac{32123}{4860} \, S_1
+\tfrac{11}{12} \, \text{ln}(z_H)^2
+\tfrac{3463}{270} \, \zeta_2
+\tfrac{197509}{8640} \, \text{ln}(z_H)
+\tfrac{31313}{240} \, S_2
\Biggr] \nonumber \\ & & 
\nonumber \\ &+&z_H^3 \, \Biggl[
-\tfrac{20245553}{64800} \, S_1
-\tfrac{141593}{1440} \, \text{ln}(z_H) \, S_1
+\tfrac{12311}{2880} \, \text{ln}(z_H)^2
+\tfrac{25951}{720} \, \zeta_2
+\tfrac{5251313}{28800} \, \text{ln}(z_H)
+\tfrac{109751}{480} \, S_2
+\tfrac{2537501}{5400}
\Biggr] \nonumber \\ & & 
\nonumber \\ &+&z_H^4 \, \Biggl[
-\tfrac{633111511}{136080} \, S_1
-\tfrac{141862}{81} \, \text{ln}(z_H) \, S_1
-\tfrac{355907}{5040} \, \zeta_2
+\tfrac{12071}{720} \, \text{ln}(z_H)^2
+\tfrac{2867831}{3360} \, S_2
+\tfrac{118206539}{37800} \, \text{ln}(z_H)
+\tfrac{1002609122909}{127008000}
\Biggr] \nonumber \\
\nonumber
\end{eqnarray}


\section{Mass difference expansion of $\Delta r^{(2)}_{\text{bos}}$ in
  the on-shell scheme}

\label{onshelldiff}

The correction $\Delta r^{(2)}_{\text{bos}}$ in the on-shell scheme
for Higgs masses in the vicinity of the $Z$ boson mass is correctly
described by an expansion in the mass difference between the Higgs
boson and the $Z$ boson and in the mass difference between the $W$ and
$Z$ bosons. The series below contains five terms in both variables
\begin{equation}
   \left( \Delta r^{(2)}_{\text{bos}} \right)^{\OS} = \left( 
    \frac{\alpha}{4 \pi \sin^2 \theta_W}\right)^2
    \sum^{4}_{n=0} \sin^{2n} \theta_W R^{\OS}_n.
\end{equation}
The transcendental numbers are the same as in the previous
section. The lack of logarithms of mass ratios follows from the fact
that a Taylor series in the mass difference does not lead to any
infrared problems. The variable $h_Z$ denotes $(M_H^2-M_Z^2)/M_Z^2$.
The first four of the coefficients were already
published in \cite{oni}
\begin{eqnarray}
\nonumber
R^{\OS}_0=
 & & \, \Biggl[
+\tfrac{20659}{48}
+62 \, \zeta_2 \, \text{ln}(3)
+\tfrac{7151}{144} \, \zeta_2
-\tfrac{403}{6} \, \zeta_3
+\tfrac{75}{4} \, S_1 \, S_2
+6 \, S_1 \, \text{ln}(3)
-\tfrac{10375}{12} \, S_1
+\tfrac{2909}{18} \, S_1^2
+\tfrac{59629}{32} \, S_2
\Biggr]
\nonumber \\ &+&h_Z \, \Biggl[
+\tfrac{2783}{288}
+\tfrac{44}{9} \, \zeta_2 \, \text{ln}(3)
-\tfrac{2489}{144} \, \zeta_2
-\tfrac{371}{18} \, \zeta_3
+\tfrac{305}{4} \, S_1 \, S_2
-\tfrac{4}{3} \, S_1 \, \text{ln}(3)
-\tfrac{959}{18} \, S_1
+\tfrac{3271}{216} \, S_1^2
+\tfrac{5201}{32} \, S_2
\Biggr]
\nonumber \\ &+&h_Z^2 \, \Biggl[
-\tfrac{43603}{1728}
-\tfrac{4}{3} \, \zeta_2 \, \text{ln}(3)
-\tfrac{4477}{864} \, \zeta_2
+\tfrac{5}{16} \, \zeta_3
+\tfrac{99}{32} \, S_1 \, S_2
+\tfrac{5}{3} \, S_1 \, \text{ln}(3)
+\tfrac{8587}{216} \, S_1
-\tfrac{823}{81} \, S_1^2
-\tfrac{3595}{192} \, S_2
\Biggr]
\nonumber \\ &+&h_Z^3 \, \Biggl[
+\tfrac{194641}{15552}
+\tfrac{4}{9} \, \zeta_2 \, \text{ln}(3)
-\tfrac{16531}{1296} \, \zeta_2
-\tfrac{35}{36} \, \zeta_3
+\tfrac{23}{8} \, S_1 \, S_2
-\tfrac{1}{3} \, S_1 \, \text{ln}(3)
-\tfrac{3491}{720} \, S_1
+\tfrac{5615}{972} \, S_1^2
-\tfrac{12373}{864} \, S_2
\Biggr]
\nonumber \\ &+&h_Z^4 \, \Biggl[
-\tfrac{88369}{17280}
+\tfrac{151487}{23328} \, \zeta_2
-\tfrac{1}{12} \, \zeta_3
+\tfrac{3}{8} \, S_1 \, S_2
+\tfrac{31}{162} \, S_1 \, \text{ln}(3)
+\tfrac{113201}{12960} \, S_1
-\tfrac{33611}{5832} \, S_1^2
-\tfrac{1937}{288} \, S_2
\Biggr] \nonumber
\end{eqnarray}
\begin{eqnarray}
R^{\OS}_1=
 & & \, \Biggl[
-\tfrac{166967}{72}
+96 \, \zeta_2 \, \text{ln}(2)
-\tfrac{2260}{9} \, \zeta_2 \, \text{ln}(3)
-\tfrac{9565}{36} \, \zeta_2
+\tfrac{4789}{18} \, \zeta_3
-\tfrac{445}{4} \, S_1 \, S_2
-\tfrac{238}{3} \, S_1 \, \text{ln}(3)
+\tfrac{31450}{9} \, S_1
-\tfrac{62045}{108} \, S_1^2
\nonumber \\ & & 
-\tfrac{45355}{8} \, S_2
\Biggr]
\nonumber \\ &+&h_Z \, \Biggl[
-\tfrac{24917}{864}
-\tfrac{68}{9} \, \zeta_2 \, \text{ln}(3)
-\tfrac{409}{432} \, \zeta_2
+\tfrac{707}{18} \, \zeta_3
-\tfrac{605}{4} \, S_1 \, S_2
+\tfrac{14}{3} \, S_1 \, \text{ln}(3)
+\tfrac{24007}{216} \, S_1
-\tfrac{8869}{324} \, S_1^2
-\tfrac{17111}{96} \, S_2
\Biggr]
\nonumber \\ &+&h_Z^2 \, \Biggl[
+\tfrac{247739}{5184}
+\tfrac{4}{3} \, \zeta_2 \, \text{ln}(3)
-\tfrac{52787}{2592} \, \zeta_2
-\tfrac{65}{16} \, \zeta_3
+\tfrac{441}{32} \, S_1 \, S_2
-S_1 \, \text{ln}(3)
-\tfrac{106343}{4320} \, S_1
+\tfrac{15737}{1296} \, S_1^2
-\tfrac{31285}{576} \, S_2
\Biggr]
\nonumber \\ &+&h_Z^3 \, \Biggl[
-\tfrac{1085639}{155520}
+\tfrac{4}{9} \, \zeta_2 \, \text{ln}(3)
+\tfrac{137425}{11664} \, \zeta_2
-\tfrac{37}{72} \, \zeta_3
+\tfrac{13}{16} \, S_1 \, S_2
+\tfrac{35}{81} \, S_1 \, \text{ln}(3)
-\tfrac{92377}{77760} \, S_1
-\tfrac{29293}{5832} \, S_1^2
+\tfrac{21419}{864} \, S_2
\Biggr]
\nonumber \\ &+&h_Z^4 \, \Biggl[
+\tfrac{4592603}{777600}
-\tfrac{532225}{23328} \, \zeta_2
-\tfrac{1}{12} \, \zeta_3
+\tfrac{3}{8} \, S_1 \, S_2
+\tfrac{7}{54} \, S_1 \, \text{ln}(3)
+\tfrac{10434679}{816480} \, S_1
+\tfrac{80429}{34992} \, S_1^2
-\tfrac{3611}{1728} \, S_2
\Biggr] \nonumber
\end{eqnarray}
\begin{eqnarray}
R^{\OS}_2=
 & & \, \Biggl[
+\tfrac{5908523}{1296}
-192 \, \zeta_2 \, \text{ln}(2)
+\tfrac{2536}{9} \, \zeta_2 \, \text{ln}(3)
+\tfrac{21139}{48} \, \zeta_2
-\tfrac{2897}{9} \, \zeta_3
+\tfrac{317}{2} \, S_1 \, S_2
+245 \, S_1 \, \text{ln}(3)
-\tfrac{1559051}{324} \, S_1
+\tfrac{5426}{9} \, S_1^2
\nonumber \\ & & 
+\tfrac{1132651}{288} \, S_2
\Biggr]
\nonumber \\ &+&h_Z \, \Biggl[
+\tfrac{746443}{15552}
-\tfrac{32}{9} \, \zeta_2 \, \text{ln}(3)
-\tfrac{10121}{324} \, \zeta_2
+\tfrac{79}{9} \, \zeta_3
-\tfrac{55}{2} \, S_1 \, S_2
+\tfrac{1}{3} \, S_1 \, \text{ln}(3)
-\tfrac{6461}{1440} \, S_1
+\tfrac{19085}{1944} \, S_1^2
-\tfrac{20417}{432} \, S_2
\Biggr]
\nonumber \\ &+&h_Z^2 \, \Biggl[
+\tfrac{214469}{155520}
+\tfrac{8}{3} \, \zeta_2 \, \text{ln}(3)
+\tfrac{43439}{2592} \, \zeta_2
-\tfrac{173}{48} \, \zeta_3
+\tfrac{231}{32} \, S_1 \, S_2
-\tfrac{41}{27} \, S_1 \, \text{ln}(3)
-\tfrac{293137}{77760} \, S_1
-\tfrac{605}{162} \, S_1^2
-\tfrac{70549}{1728} \, S_2
\Biggr]
\nonumber \\ &+&h_Z^3 \, \Biggl[
-\tfrac{3867907}{1166400}
-\tfrac{13853}{486} \, \zeta_2
+\tfrac{9}{8} \, \zeta_3
-\tfrac{81}{16} \, S_1 \, S_2
+\tfrac{83}{81} \, S_1 \, \text{ln}(3)
+\tfrac{8558191}{489888} \, S_1
+\tfrac{311}{729} \, S_1^2
+\tfrac{162193}{2592} \, S_2
\Biggr]
\nonumber \\ &+&h_Z^4 \, \Biggl[
+\tfrac{7035509}{979776}
-\tfrac{16024969}{559872} \, \zeta_2
+\tfrac{1}{12} \, \zeta_3
-\tfrac{3}{8} \, S_1 \, S_2
+\tfrac{1}{27} \, S_1 \, \text{ln}(3)
+\tfrac{21579127}{699840} \, S_1
-\tfrac{58837}{13122} \, S_1^2
-\tfrac{41531}{7776} \, S_2
\Biggr] \nonumber
\end{eqnarray}
\begin{eqnarray}
R^{\OS}_3=
 & & \, \Biggl[
-\tfrac{5582263}{1944}
-\tfrac{704}{9} \, \zeta_2 \, \text{ln}(3)
-\tfrac{281321}{648} \, \zeta_2
+\tfrac{5581}{36} \, \zeta_3
-\tfrac{637}{8} \, S_1 \, S_2
-\tfrac{586}{3} \, S_1 \, \text{ln}(3)
+\tfrac{109883}{40} \, S_1
-\tfrac{17078}{81} \, S_1^2
-\tfrac{521903}{432} \, S_2
\Biggr]
\nonumber \\ &+&h_Z \, \Biggl[
+\tfrac{437711}{25920}
-\tfrac{8}{9} \, \zeta_2 \, \text{ln}(3)
+\tfrac{139991}{5832} \, \zeta_2
+\tfrac{343}{36} \, \zeta_3
-\tfrac{319}{8} \, S_1 \, S_2
-\tfrac{181}{81} \, S_1 \, \text{ln}(3)
-\tfrac{536677}{38880} \, S_1
-\tfrac{6535}{5832} \, S_1^2
-\tfrac{659}{12} \, S_2
\Biggr]
\nonumber \\ &+&h_Z^2 \, \Biggl[
-\tfrac{15188507}{777600}
+\tfrac{8}{3} \, \zeta_2 \, \text{ln}(3)
-\tfrac{355457}{46656} \, \zeta_2
-\tfrac{29}{24} \, \zeta_3
-\tfrac{57}{16} \, S_1 \, S_2
-\tfrac{125}{81} \, S_1 \, \text{ln}(3)
+\tfrac{27541403}{1088640} \, S_1
-\tfrac{45707}{8748} \, S_1^2
+\tfrac{69397}{3456} \, S_2
\Biggr]
\nonumber \\ &+&h_Z^3 \, \Biggl[
-\tfrac{25783}{2449440}
-\tfrac{8}{9} \, \zeta_2 \, \text{ln}(3)
-\tfrac{926609}{17496} \, \zeta_2
+\tfrac{163}{36} \, \zeta_3
-\tfrac{139}{8} \, S_1 \, S_2
+\tfrac{299}{243} \, S_1 \, \text{ln}(3)
+\tfrac{4479889}{136080} \, S_1
-\tfrac{11707}{13122} \, S_1^2
+\tfrac{1926995}{15552} \, S_2
\Biggr]
\nonumber \\ &+&h_Z^4 \, \Biggl[
+\tfrac{9642347291}{342921600}
-\tfrac{559304957}{5038848} \, \zeta_2
+\tfrac{1}{2} \, \zeta_3
-\tfrac{9}{4} \, S_1 \, S_2
-\tfrac{401}{2187} \, S_1 \, \text{ln}(3)
+\tfrac{386672993}{4199040} \, S_1
-\tfrac{646097}{157464} \, S_1^2
+\tfrac{27433}{3888} \, S_2
\Biggr] \nonumber
\end{eqnarray}
\begin{eqnarray}
R^{\OS}_4=
 & & \, \Biggl[
+\tfrac{23299549}{38880}
+\tfrac{376}{9} \, \zeta_2 \, \text{ln}(3)
+\tfrac{4413259}{11664} \, \zeta_2
-\tfrac{343}{18} \, \zeta_3
-\tfrac{509}{4} \, S_1 \, S_2
+\tfrac{6223}{162} \, S_1 \, \text{ln}(3)
-\tfrac{25615403}{38880} \, S_1
-\tfrac{12640}{243} \, S_1^2
+\tfrac{1314829}{4320} \, S_2
\Biggr]
\nonumber \\ &+&h_Z \, \Biggl[
-\tfrac{50017}{5400}
+\tfrac{4}{9} \, \zeta_2 \, \text{ln}(3)
+\tfrac{1726547}{58320} \, \zeta_2
+\tfrac{91}{9} \, \zeta_3
-47 \, S_1 \, S_2
-\tfrac{73}{18} \, S_1 \, \text{ln}(3)
-\tfrac{2747231}{816480} \, S_1
-\tfrac{114823}{17496} \, S_1^2
+\tfrac{31919}{1440} \, S_2
\Biggr]
\nonumber \\ &+&h_Z^2 \, \Biggl[
-\tfrac{637699471}{16329600}
+\tfrac{4}{3} \, \zeta_2 \, \text{ln}(3)
-\tfrac{3912127}{174960} \, \zeta_2
+\tfrac{239}{48} \, \zeta_3
-\tfrac{861}{32} \, S_1 \, S_2
-\tfrac{293}{162} \, S_1 \, \text{ln}(3)
+\tfrac{2714011}{68040} \, S_1
-\tfrac{530147}{104976} \, S_1^2
+\tfrac{284597}{2592} \, S_2
\Biggr]
\nonumber \\ &+&h_Z^3 \, \Biggl[
+\tfrac{308051815}{13716864}
-\tfrac{20}{9} \, \zeta_2 \, \text{ln}(3)
-\tfrac{3545221483}{25194240} \, \zeta_2
+\tfrac{91}{9} \, \zeta_3
-38 \, S_1 \, S_2
+\tfrac{3365}{4374} \, S_1 \, \text{ln}(3)
+\tfrac{1296774019}{14696640} \, S_1
+\tfrac{40363}{78732} \, S_1^2 \nonumber \\ & &
+\tfrac{16604743}{77760} \, S_2
\Biggr]
\nonumber \\ &+&h_Z^4 \, \Biggl[
+\tfrac{368208335437}{4115059200}
-\tfrac{8071459297}{26873856} \, \zeta_2
+\tfrac{5}{4} \, \zeta_3
-\tfrac{45}{8} \, S_1 \, S_2
-\tfrac{2627}{4374} \, S_1 \, \text{ln}(3)
+\tfrac{6605645449}{29393280} \, S_1
-\tfrac{2503867}{944784} \, S_1^2
+\tfrac{3650279}{116640} \, S_2
\Biggr] \nonumber
\end{eqnarray}


\section{Large Higgs boson mass expansion of $\Delta
 r^{(2)}_{\text{bos}}$ in the \MSb scheme}

\label{msbar}

In this appendix, $\Delta r^{(2)}_{\text{bos}}$ renormalised in the
\MSb scheme is presented as a twofold expansion in the large Higgs
boson mass and in the mass difference between the $W$ and the $Z$
boson. The expansion is parametrised as follows
\begin{equation}
  \left( \Delta r^{(2)}_{\text{bos}} \right)^{\msb} =
  \left( \frac{\alpha}{4 \pi \sin^2 \theta_W} \right)^2 \frac{m_H^4}{m_Z^4}
  \sum_{n=0}^{5} \sin^{2n} \theta_W R^{\msb}_n.
\end{equation}
The parameters, {\it i.e.} masses and the coupling constant are
in the \MSb scheme. Apart from the numbers
Eq.~\ref{transcendental}, it is assumed that $\text{ln}(m_{Z,H}^2) = 
\text{ln}(m_{Z,H}^2/\mu^2)$, $\mu$ being the \MSb
renormalisation scale.

\begin{eqnarray}
\nonumber
R^{\msb}_0 &=&
\Biggl[ -\tfrac{243}{32} \, S_2
-\tfrac{27}{8} \, \text{ln}(m_H^2)
+\tfrac{1}{4} \, \zeta_2
+\tfrac{27}{32} \, \text{ln}(m_H^2)^2
+\tfrac{457}{128}
\Biggr] \nonumber \\
&+&z_H \, \Biggl[
-\tfrac{243}{32} \, S_2
-\tfrac{237}{32}
-\tfrac{45}{16} \, \zeta_2
-\tfrac{9}{4} \, \text{ln}(m_H^2)^2
-\tfrac{9}{8} \, \text{ln}(m_Z^2) \, \text{ln}(m_H^2)
+\tfrac{21}{16} \, \text{ln}(m_Z^2)
+\tfrac{123}{16} \, \text{ln}(m_H^2)
\Biggr] \nonumber \\ & & 
\nonumber \\ &+&z_H^2 \, \Biggl[
-\tfrac{567}{32} \, S_2
-\tfrac{53}{3} \, \text{ln}(m_H^2)
-\tfrac{805}{48} \, \text{ln}(m_Z^2)
+\tfrac{195}{32} \, \text{ln}(m_Z^2)^2
+\tfrac{131}{16} \, \text{ln}(m_H^2)^2
+\tfrac{113}{8} \, \zeta_2
+\tfrac{34243}{1152}
-2 \, \text{ln}(m_Z^2) \, \text{ln}(m_H^2)
\Biggr] \nonumber \\ & & 
\nonumber \\ &+&z_H^3 \, \Biggl[
-\tfrac{55831}{576}
-\tfrac{2881}{48} \, \text{ln}(m_H^2)
-\tfrac{649}{16} \, \text{ln}(m_Z^2)^2
-\tfrac{15}{16} \, \text{ln}(m_H^2)^2
+\tfrac{3}{16} \, \zeta_2
+\tfrac{37}{4} \, \text{ln}(m_Z^2) \, \text{ln}(m_H^2)
+\tfrac{4697}{24} \, \text{ln}(m_Z^2)
+\tfrac{14499}{32} \, S_2
\Biggr] \nonumber \\ & & 
\nonumber \\ &+&z_H^4 \, \Biggl[
-\tfrac{7783}{96} \, \text{ln}(m_H^2)
-\tfrac{731}{32} \, \text{ln}(m_Z^2)^2
-\tfrac{307}{16} \, \zeta_2
-\tfrac{287}{32} \, \text{ln}(m_H^2)^2
+\tfrac{151399}{14400}
+\tfrac{509}{16} \, \text{ln}(m_Z^2) \, \text{ln}(m_H^2)
+\tfrac{5215}{96} \, \text{ln}(m_Z^2)
+243 \, S_2
\Biggr] \nonumber \\ & & 
\nonumber \\ &+&z_H^5 \, \Biggl[
-\tfrac{68129}{672} \, \text{ln}(m_H^2)
-\tfrac{931}{16} \, \zeta_2
-\tfrac{779}{16} \, \text{ln}(m_Z^2)^2
-\tfrac{453}{16} \, \text{ln}(m_H^2)^2
+\tfrac{44394899}{1411200}
+\tfrac{63761}{672} \, \text{ln}(m_Z^2)
+\tfrac{11745}{32} \, S_2
\nonumber \\ && \,\,
+77 \, \text{ln}(m_Z^2) \, \text{ln}(m_H^2)
\Biggr] \nonumber \\ 
R^{\msb}_1&=&
\Biggl[ -\tfrac{243}{16} \, S_2
-\tfrac{27}{4} \, \text{ln}(m_H^2)
+\tfrac{1}{2} \, \zeta_2
+\tfrac{27}{16} \, \text{ln}(m_H^2)^2
+\tfrac{457}{64}
\Biggr] \nonumber \\
&+&z_H \, \Biggl[
-\tfrac{181}{16}
-\tfrac{243}{32} \, S_2
-\tfrac{63}{16} \, \zeta_2
-\tfrac{45}{16} \, \text{ln}(m_H^2)^2
-\tfrac{27}{16} \, \text{ln}(m_Z^2) \, \text{ln}(m_H^2)
+\tfrac{63}{32} \, \text{ln}(m_Z^2)
+\tfrac{177}{16} \, \text{ln}(m_H^2)
\Biggr] \nonumber \\ & & 
\nonumber \\ &+&z_H^2 \, \Biggl[
-\tfrac{1269}{32} \, S_2
-\tfrac{1073}{96} \, \text{ln}(m_Z^2)
-\tfrac{113}{48} \, \text{ln}(m_H^2)
-\tfrac{5}{8} \, \text{ln}(m_Z^2) \, \text{ln}(m_H^2)
+\tfrac{19}{32} \, \text{ln}(m_Z^2)^2
+\tfrac{43}{16} \, \zeta_2
+\tfrac{91}{32} \, \text{ln}(m_H^2)^2
+\tfrac{20845}{576}
\Biggr] \nonumber \\ & & 
\nonumber \\ &+&z_H^3 \, \Biggl[
-\tfrac{4833}{8} \, S_2
-\tfrac{10445}{96} \, \text{ln}(m_Z^2)
-\tfrac{54875}{1152}
-\tfrac{167}{32} \, \text{ln}(m_H^2)^2
+\tfrac{1}{4} \, \text{ln}(m_Z^2) \, \text{ln}(m_H^2)
+\tfrac{847}{32} \, \text{ln}(m_Z^2)^2
+\tfrac{3835}{96} \, \text{ln}(m_H^2)
-12 \, \zeta_2
\Biggr] \nonumber \\ & & 
\nonumber \\ &+&z_H^4 \, \Biggl[
-\tfrac{16659}{32} \, S_2
-\tfrac{3929}{40} \, \text{ln}(m_Z^2)
-\tfrac{6123}{160}
+\tfrac{297}{32} \, \text{ln}(m_H^2)^2
+\tfrac{329}{16} \, \zeta_2
+\tfrac{1239}{32} \, \text{ln}(m_Z^2)^2
+\tfrac{5499}{40} \, \text{ln}(m_H^2)
-48 \, \text{ln}(m_Z^2) \, \text{ln}(m_H^2)
\Biggr] \nonumber \\ & & 
\nonumber \\ &+&z_H^5 \, \Biggl[
-\tfrac{9045}{8} \, S_2
-\tfrac{168283}{840} \, \text{ln}(m_Z^2)
-\tfrac{1579}{8} \, \text{ln}(m_Z^2) \, \text{ln}(m_H^2)
-\tfrac{84807787}{705600}
+\tfrac{2171}{32} \, \text{ln}(m_H^2)^2
+\tfrac{4145}{32} \, \text{ln}(m_Z^2)^2
+\tfrac{1119}{8} \, \zeta_2
\nonumber \\ && 
+\tfrac{200833}{840} \, \text{ln}(m_H^2)
\Biggr] \nonumber \\ 
R^{\msb}_2&=&
 \Biggl[-\tfrac{729}{32} \, S_2
-\tfrac{81}{8} \, \text{ln}(m_H^2)
+\tfrac{3}{4} \, \zeta_2
+\tfrac{81}{32} \, \text{ln}(m_H^2)^2
+\tfrac{1371}{128}
\Biggr] \nonumber \\ 
&+&z_H \, \Biggl[
-\tfrac{473}{32}
-\tfrac{243}{32} \, S_2
-\tfrac{81}{16} \, \zeta_2
-\tfrac{27}{8} \, \text{ln}(m_H^2)^2
-\tfrac{9}{4} \, \text{ln}(m_Z^2) \, \text{ln}(m_H^2)
+\tfrac{21}{8} \, \text{ln}(m_Z^2)
+\tfrac{225}{16} \, \text{ln}(m_H^2)
\Biggr] \nonumber \\ & & 
\nonumber \\ &+&z_H^2 \, \Biggl[
-\tfrac{529}{48} \, \text{ln}(m_Z^2)
-\tfrac{295}{96} \, \text{ln}(m_H^2)
-\tfrac{45}{16} \, \text{ln}(m_Z^2) \, \text{ln}(m_H^2)
-\tfrac{391}{384}
+\tfrac{43}{32} \, \text{ln}(m_Z^2)^2
+\tfrac{81}{16} \, \text{ln}(m_H^2)^2
+\tfrac{127}{16} \, \zeta_2
+114 \, S_2
\Biggr] \nonumber \\ & & 
\nonumber \\ &+&z_H^3 \, \Biggl[
-\tfrac{1759}{96} \, \text{ln}(m_Z^2)
-\tfrac{1213}{96} \, \text{ln}(m_H^2)
-\tfrac{17}{4} \, \text{ln}(m_H^2)^2
-\tfrac{23}{8} \, \text{ln}(m_Z^2)^2
+\tfrac{59}{8} \, \text{ln}(m_Z^2) \, \text{ln}(m_H^2)
+\tfrac{38687}{1152}
+\tfrac{987}{16} \, S_2
-7 \, \zeta_2
\Biggr] \nonumber \\ & & 
\nonumber \\ &+&z_H^4 \, \Biggl[
-\tfrac{16447}{480} \, \text{ln}(m_H^2)
-\tfrac{257}{8} \, \zeta_2
-\tfrac{63}{2} \, \text{ln}(m_Z^2)^2
-\tfrac{255}{16} \, \text{ln}(m_H^2)^2
-\tfrac{5033}{480} \, \text{ln}(m_Z^2)
+\tfrac{759}{16} \, \text{ln}(m_Z^2) \, \text{ln}(m_H^2)
+\tfrac{1948279}{28800}
+\tfrac{11559}{32} \, S_2
\Biggr] \nonumber \\ & & 
\nonumber \\ &+&z_H^5 \, \Biggl[
-\tfrac{3573}{16} \, \zeta_2
-\tfrac{2775}{16} \, \text{ln}(m_Z^2)^2
-\tfrac{1761}{16} \, \text{ln}(m_H^2)^2
-\tfrac{142399}{3360} \, \text{ln}(m_H^2)
-\tfrac{92381}{3360} \, \text{ln}(m_Z^2)
+\tfrac{108755447}{470400}
+\tfrac{567}{2} \, \text{ln}(m_Z^2) \, \text{ln}(m_H^2)
\nonumber \\ &&
+\tfrac{40827}{32} \, S_2
\Biggr] \nonumber
\nonumber \\
R^{\msb}_3 &=&
\Biggl[-\tfrac{243}{8} \, S_2
-\tfrac{27}{2} \, \text{ln}(m_H^2)
+\tfrac{27}{8} \, \text{ln}(m_H^2)^2
+\tfrac{457}{32}
+\zeta_2
\Biggr] \nonumber \\ 
&+&z_H \, \Biggl[
-\tfrac{2315}{128}
-\tfrac{243}{32} \, S_2
-\tfrac{99}{16} \, \zeta_2
-\tfrac{63}{16} \, \text{ln}(m_H^2)^2
-\tfrac{45}{16} \, \text{ln}(m_Z^2) \, \text{ln}(m_H^2)
+\tfrac{105}{32} \, \text{ln}(m_Z^2)
+\tfrac{1083}{64} \, \text{ln}(m_H^2)
\Biggr] \nonumber \\ & & 
\nonumber \\ &+&z_H^2 \, \Biggl[
-\tfrac{2161}{192} \, \text{ln}(m_Z^2)
-\tfrac{179}{48} \, \text{ln}(m_H^2)
+\tfrac{67}{32} \, \text{ln}(m_Z^2)^2
+\tfrac{233}{32} \, \text{ln}(m_H^2)^2
+\tfrac{333}{32} \, S_2
+\tfrac{257}{16} \, \zeta_2
+\tfrac{25579}{1152}
-5 \, \text{ln}(m_Z^2) \, \text{ln}(m_H^2)
\Biggr] \nonumber \\ & & 
\nonumber \\ &+&z_H^3 \, \Biggl[
-\tfrac{629}{32} \, \text{ln}(m_H^2)
-\tfrac{375}{32} \, S_2
-\tfrac{113}{16} \, \zeta_2
-\tfrac{145}{32} \, \text{ln}(m_H^2)^2
-\tfrac{93}{32} \, \text{ln}(m_Z^2)^2
+\tfrac{171}{32} \, \text{ln}(m_Z^2)
+\tfrac{127}{16} \, \text{ln}(m_Z^2) \, \text{ln}(m_H^2)
+\tfrac{13183}{576}
\Biggr] \nonumber \\ & & 
\nonumber \\ &+&z_H^4 \, \Biggl[
-\tfrac{657}{16} \, S_2
-\tfrac{7789}{480} \, \text{ln}(m_H^2)
-\tfrac{11}{4} \, \text{ln}(m_Z^2)^2
-\tfrac{11}{4} \, \text{ln}(m_H^2)^2
+\tfrac{11}{2} \, \text{ln}(m_Z^2) \, \text{ln}(m_H^2)
+\tfrac{3047}{240} \, \text{ln}(m_Z^2)
+\tfrac{497737}{28800}
-6 \, \zeta_2
\Biggr] \nonumber \\ & & 
\nonumber \\ &+&z_H^5 \, \Biggl[
-\tfrac{17787}{32} \, S_2
-\tfrac{1571}{16} \, \text{ln}(m_Z^2) \, \text{ln}(m_H^2)
-\tfrac{97161}{1120} \, \text{ln}(m_H^2)
-\tfrac{41295691}{1411200}
+\tfrac{609}{16} \, \text{ln}(m_H^2)^2
+\tfrac{481}{8} \, \text{ln}(m_Z^2)^2
+\tfrac{1219}{16} \, \zeta_2
\nonumber \\ && 
+\tfrac{55681}{420} \, \text{ln}(m_Z^2)
\Biggr] \nonumber 
\nonumber \\
R^{\msb}_4 &=&
\Biggl[-\tfrac{1215}{32} \, S_2
-\tfrac{135}{8} \, \text{ln}(m_H^2)
+\tfrac{5}{4} \, \zeta_2
+\tfrac{135}{32} \, \text{ln}(m_H^2)^2
+\tfrac{2285}{128}
\Biggr] \nonumber \\ 
&+&z_H \, \Biggl[
-\tfrac{6817}{320}
-\tfrac{243}{32} \, S_2
-\tfrac{117}{16} \, \zeta_2
-\tfrac{9}{2} \, \text{ln}(m_H^2)^2
-\tfrac{27}{8} \, \text{ln}(m_Z^2) \, \text{ln}(m_H^2)
+\tfrac{63}{16} \, \text{ln}(m_Z^2)
+\tfrac{3153}{160} \, \text{ln}(m_H^2)
\Biggr] \nonumber \\ & & 
\nonumber \\ &+&z_H^2 \, \Biggl[
-\tfrac{1347}{32} \, S_2
-\tfrac{349}{30} \, \text{ln}(m_Z^2)
-\tfrac{115}{16} \, \text{ln}(m_Z^2) \, \text{ln}(m_H^2)
-\tfrac{851}{192} \, \text{ln}(m_H^2)
+\tfrac{91}{32} \, \text{ln}(m_Z^2)^2
+\tfrac{19}{2} \, \text{ln}(m_H^2)^2
+\tfrac{171}{8} \, \zeta_2
+\tfrac{51769}{1440}
\Biggr] \nonumber \\ & & 
\nonumber \\ &+&z_H^3 \, \Biggl[
-\tfrac{1243}{48} \, \text{ln}(m_H^2)
-\tfrac{55}{4} \, S_2
-\tfrac{53}{8} \, \zeta_2
-\tfrac{77}{16} \, \text{ln}(m_H^2)^2
-\tfrac{47}{16} \, \text{ln}(m_Z^2)^2
+\tfrac{311}{40}
+\tfrac{17}{2} \, \text{ln}(m_Z^2) \, \text{ln}(m_H^2)
+\tfrac{883}{60} \, \text{ln}(m_Z^2)
\Biggr] \nonumber \\ & & 
\nonumber \\ &+&z_H^4 \, \Biggl[
-\tfrac{2897}{192} \, \text{ln}(m_H^2)
-\tfrac{15}{16} \, \zeta_2
-\tfrac{17}{32} \, S_2
-\tfrac{15}{32} \, \text{ln}(m_Z^2)^2
-\tfrac{15}{32} \, \text{ln}(m_H^2)^2
+\tfrac{15}{16} \, \text{ln}(m_Z^2) \, \text{ln}(m_H^2)
+\tfrac{38587}{14400}
+\tfrac{1013}{192} \, \text{ln}(m_Z^2)
\Biggr] \nonumber \\ & & 
\nonumber \\ &+&z_H^5 \, \Biggl[
-\tfrac{1366523}{33600}
-\tfrac{179}{8} \, \text{ln}(m_Z^2) \, \text{ln}(m_H^2)
-\tfrac{3497}{160} \, \text{ln}(m_H^2)
+\tfrac{179}{16} \, \text{ln}(m_Z^2)^2
+\tfrac{179}{16} \, \text{ln}(m_H^2)^2
+\tfrac{1281}{80} \, \text{ln}(m_Z^2)
+\tfrac{183}{8} \, \zeta_2
+\tfrac{791}{16} \, S_2
\Biggr] \nonumber 
\nonumber \\
R^{\msb}_5&=&
 \Biggl[-\tfrac{729}{16} \, S_2
-\tfrac{81}{4} \, \text{ln}(m_H^2)
+\tfrac{3}{2} \, \zeta_2
+\tfrac{81}{16} \, \text{ln}(m_H^2)^2
+\tfrac{1371}{64}
\Biggr] \nonumber \\ 
&+&z_H \, \Biggl[
-\tfrac{7829}{320}
-\tfrac{135}{16} \, \zeta_2
-\tfrac{243}{32} \, S_2
-\tfrac{81}{16} \, \text{ln}(m_H^2)^2
-\tfrac{63}{16} \, \text{ln}(m_Z^2) \, \text{ln}(m_H^2)
+\tfrac{147}{32} \, \text{ln}(m_Z^2)
+\tfrac{3591}{160} \, \text{ln}(m_H^2)
\Biggr] \nonumber \\ & & 
\nonumber \\ &+&z_H^2 \, \Biggl[
-\tfrac{24949}{288} \, S_2
-\tfrac{1931}{160} \, \text{ln}(m_Z^2)
-\tfrac{75}{8} \, \text{ln}(m_Z^2) \, \text{ln}(m_H^2)
-\tfrac{497}{96} \, \text{ln}(m_H^2)
+\tfrac{115}{32} \, \text{ln}(m_Z^2)^2
+\tfrac{375}{32} \, \text{ln}(m_H^2)^2
+\tfrac{419}{16} \, \zeta_2
+\tfrac{12475499}{259200}
\Biggr] \nonumber \\ & & 
\nonumber \\ &+&z_H^3 \, \Biggl[
-\tfrac{641}{20} \, \text{ln}(m_H^2)
-\tfrac{3211}{288} \, S_2
-\tfrac{99}{16} \, \zeta_2
-\tfrac{163}{32} \, \text{ln}(m_H^2)^2
-\tfrac{34321}{10368}
-\tfrac{95}{32} \, \text{ln}(m_Z^2)^2
+\tfrac{145}{16} \, \text{ln}(m_Z^2) \, \text{ln}(m_H^2)
+\tfrac{20329}{960} \, \text{ln}(m_Z^2)
\Biggr] \nonumber \\ & & 
\nonumber \\ &+&z_H^4 \, \Biggl[
-\tfrac{4799}{240} \, \text{ln}(m_H^2)
-\tfrac{29}{8} \, \text{ln}(m_Z^2) \, \text{ln}(m_H^2)
+\tfrac{113273}{259200}
+\tfrac{29}{16} \, \text{ln}(m_Z^2)^2
+\tfrac{29}{16} \, \text{ln}(m_H^2)^2
+\tfrac{343}{144} \, S_2
+\tfrac{29}{8} \, \zeta_2
+\tfrac{1363}{192} \, \text{ln}(m_Z^2)
\Biggr] \nonumber \\ & & 
\nonumber \\ &+&z_H^5 \, \Biggl[
-\tfrac{57473}{960} \, \text{ln}(m_H^2)
-\tfrac{321}{8} \, \text{ln}(m_Z^2) \, \text{ln}(m_H^2)
-\tfrac{39526351}{1814400}
+\tfrac{97}{36} \, S_2
+\tfrac{321}{16} \, \text{ln}(m_Z^2)^2
+\tfrac{321}{16} \, \text{ln}(m_H^2)^2
+\tfrac{321}{8} \, \zeta_2
+\tfrac{5627}{96} \, \text{ln}(m_Z^2)
\Biggr] \nonumber
\end{eqnarray}


\section{Mass difference expansion of $\Delta r^{(2)}_{\text{bos}}$ in
  the \MSb scheme}

The correction in the \MSb scheme is given by six coefficients in the
double expansion in the mass differences between the $W$ and $Z$
bosons and between the Higgs boson and the $Z$ bosons
\begin{equation}
  \left( \Delta r^{(2)}_{\text{bos}} \right)^{\msb} =
  \left( \frac{\alpha}{4 \pi \sin^2 \theta_W} \right)^2
  \sum_{n=0}^{5} \sin^{2n} \theta_W R^{\msb}_n.
\end{equation}
All parameters are in the \MSb scheme and $h_Z =
(m_H^2-m_Z^2)/m_Z^2$. Note also that the logaritms contain the
renormalisation scale as $\text{ln}(m_Z^2)  = \text{ln}(m_Z^2/\mu^2)$.
\begin{eqnarray}
\nonumber
R^{\msb}_0=
 & & \, \Biggl[
-\tfrac{3673}{48}
+\tfrac{15}{16} \, \zeta_2
+\tfrac{7749}{32} \, S_2
+\tfrac{389}{4} \, \text{ln}(m_Z^2)
-\tfrac{45}{2} \, \text{ln}(m_Z^2)^2
\Biggr]
\nonumber \\ &+&h_Z \, \Biggl[
+\tfrac{37739}{576}
-\tfrac{23}{4} \, \zeta_2
-\tfrac{4053}{16} \, S_2
-\tfrac{4345}{48} \, \text{ln}(m_Z^2)
+\tfrac{489}{16} \, \text{ln}(m_Z^2)^2
\Biggr]
\nonumber \\ &+&h_Z^2 \, \Biggl[
-\tfrac{66773}{1152}
+\tfrac{7}{4} \, \zeta_2
+\tfrac{9083}{32} \, S_2
+\tfrac{6857}{96} \, \text{ln}(m_Z^2)
-\tfrac{1005}{32} \, \text{ln}(m_Z^2)^2
\Biggr]
\nonumber \\ &+&h_Z^3 \, \Biggl[
+\tfrac{93671}{1728}
-3 \, \zeta_2
-\tfrac{14509}{48} \, S_2
-\tfrac{26611}{480} \, \text{ln}(m_Z^2)
+\tfrac{129}{4} \, \text{ln}(m_Z^2)^2
\Biggr]
\nonumber \\ &+&h_Z^4 \, \Biggl[
-\tfrac{829957}{17280}
+3 \, \zeta_2
+\tfrac{7321}{24} \, S_2
+\tfrac{35641}{960} \, \text{ln}(m_Z^2)
-\tfrac{129}{4} \, \text{ln}(m_Z^2)^2
\Biggr]
\nonumber \\ &+&h_Z^5 \, \Biggl[
+\tfrac{7597621}{181440}
-3 \, \zeta_2
-\tfrac{44257}{144} \, S_2
-\tfrac{29863}{1680} \, \text{ln}(m_Z^2)
+\tfrac{129}{4} \, \text{ln}(m_Z^2)^2
\Biggr] \nonumber
\end{eqnarray}
\begin{eqnarray}
R^{\msb}_1=
 & & \, \Biggl[
+\tfrac{1325}{36}
-\tfrac{17}{4} \, \zeta_2
-\tfrac{4119}{8} \, S_2
-\tfrac{321}{4} \, \text{ln}(m_Z^2)
+\tfrac{43}{2} \, \text{ln}(m_Z^2)^2
\Biggr]
\nonumber \\ &+&h_Z \, \Biggl[
-\tfrac{1333}{96}
+\tfrac{9}{4} \, \zeta_2
+\tfrac{1685}{4} \, S_2
+\tfrac{915}{16} \, \text{ln}(m_Z^2)
-\tfrac{181}{8} \, \text{ln}(m_Z^2)^2
\Biggr]
\nonumber \\ &+&h_Z^2 \, \Biggl[
+\tfrac{18011}{576}
-2 \, \zeta_2
-\tfrac{1929}{4} \, S_2
-\tfrac{4013}{80} \, \text{ln}(m_Z^2)
+\tfrac{371}{16} \, \text{ln}(m_Z^2)^2
\Biggr]
\nonumber \\ &+&h_Z^3 \, \Biggl[
-\tfrac{630583}{17280}
+2 \, \zeta_2
+\tfrac{2713}{6} \, S_2
+\tfrac{31199}{960} \, \text{ln}(m_Z^2)
-\tfrac{43}{2} \, \text{ln}(m_Z^2)^2
\Biggr]
\nonumber \\ &+&h_Z^4 \, \Biggl[
+\tfrac{79808347}{1814400}
-2 \, \zeta_2
-\tfrac{64831}{144} \, S_2
-\tfrac{56941}{3360} \, \text{ln}(m_Z^2)
+\tfrac{43}{2} \, \text{ln}(m_Z^2)^2
\Biggr]
\nonumber \\ &+&h_Z^5 \, \Biggl[
-\tfrac{95205863}{1814400}
+2 \, \zeta_2
+\tfrac{16235}{36} \, S_2
+\tfrac{2561}{960} \, \text{ln}(m_Z^2)
-\tfrac{43}{2} \, \text{ln}(m_Z^2)^2
\Biggr] \nonumber
\end{eqnarray}
\begin{eqnarray}
R^{\msb}_2=
 & & \, \Biggl[
+\tfrac{1519}{72}
-\tfrac{3}{16} \, \zeta_2
+\tfrac{1671}{32} \, S_2
-\tfrac{607}{12} \, \text{ln}(m_Z^2)
+\tfrac{3}{4} \, \text{ln}(m_Z^2)^2
\Biggr]
\nonumber \\ &+&h_Z \, \Biggl[
-\tfrac{21679}{864}
+\tfrac{3}{4} \, \zeta_2
-\tfrac{541}{24} \, S_2
+\tfrac{27307}{480} \, \text{ln}(m_Z^2)
-\tfrac{13}{16} \, \text{ln}(m_Z^2)^2
\Biggr]
\nonumber \\ &+&h_Z^2 \, \Biggl[
+\tfrac{889057}{25920}
+\tfrac{7}{4} \, \zeta_2
-\tfrac{20549}{288} \, S_2
-\tfrac{79529}{960} \, \text{ln}(m_Z^2)
+\tfrac{89}{32} \, \text{ln}(m_Z^2)^2
\Biggr]
\nonumber \\ &+&h_Z^3 \, \Biggl[
-\tfrac{11857673}{453600}
-\zeta_2
+\tfrac{1073}{36} \, S_2
+\tfrac{313069}{3360} \, \text{ln}(m_Z^2)
-\tfrac{1}{4} \, \text{ln}(m_Z^2)^2
\Biggr]
\nonumber \\ &+&h_Z^4 \, \Biggl[
+\tfrac{4154501}{272160}
+\zeta_2
-\tfrac{5905}{432} \, S_2
-\tfrac{356431}{3360} \, \text{ln}(m_Z^2)
+\tfrac{1}{4} \, \text{ln}(m_Z^2)^2
\Biggr]
\nonumber \\ &+&h_Z^5 \, \Biggl[
-\tfrac{1060729937}{114307200}
-\zeta_2
+\tfrac{10309}{648} \, S_2
+\tfrac{270507}{2240} \, \text{ln}(m_Z^2)
-\tfrac{1}{4} \, \text{ln}(m_Z^2)^2
\Biggr] \nonumber
\end{eqnarray}
\begin{eqnarray}
R^{\msb}_3=
 & & \, \Biggl[
+\tfrac{5707}{108}
+\tfrac{21}{4} \, \zeta_2
-\tfrac{1325}{12} \, S_2
-\tfrac{1223}{40} \, \text{ln}(m_Z^2)
+\tfrac{3}{2} \, \text{ln}(m_Z^2)^2
\Biggr]
\nonumber \\ &+&h_Z \, \Biggl[
-\tfrac{1230667}{25920}
+\tfrac{5}{4} \, \zeta_2
+\tfrac{265}{36} \, S_2
+\tfrac{13787}{480} \, \text{ln}(m_Z^2)
-\tfrac{1}{2} \, \text{ln}(m_Z^2)^2
\Biggr]
\nonumber \\ &+&h_Z^2 \, \Biggl[
+\tfrac{66550571}{907200}
+\tfrac{7}{2} \, \zeta_2
-\tfrac{17893}{144} \, S_2
-\tfrac{86971}{1680} \, \text{ln}(m_Z^2)
+\tfrac{31}{8} \, \text{ln}(m_Z^2)^2
\Biggr]
\nonumber \\ &+&h_Z^3 \, \Biggl[
-\tfrac{13312007}{155520}
-2 \, \zeta_2
+\tfrac{40147}{432} \, S_2
+\tfrac{5683}{105} \, \text{ln}(m_Z^2)
-\tfrac{1}{2} \, \text{ln}(m_Z^2)^2
\Biggr]
\nonumber \\ &+&h_Z^4 \, \Biggl[
+\tfrac{2337399703}{28576800}
+2 \, \zeta_2
-\tfrac{33491}{648} \, S_2
-\tfrac{42847}{720} \, \text{ln}(m_Z^2)
+\tfrac{1}{2} \, \text{ln}(m_Z^2)^2
\Biggr]
\nonumber \\ &+&h_Z^5 \, \Biggl[
-\tfrac{374331501521}{4115059200}
-2 \, \zeta_2
+\tfrac{43928}{729} \, S_2
+\tfrac{2708207}{40320} \, \text{ln}(m_Z^2)
-\tfrac{1}{2} \, \text{ln}(m_Z^2)^2
\Biggr] \nonumber
\end{eqnarray}
\begin{eqnarray}
R^{\msb}_4=
 & & \, \Biggl[
+\tfrac{247637}{4320}
+\tfrac{63}{8} \, \zeta_2
-\tfrac{8731}{48} \, S_2
-\tfrac{2503}{80} \, \text{ln}(m_Z^2)
+\tfrac{9}{4} \, \text{ln}(m_Z^2)^2
\Biggr]
\nonumber \\ &+&h_Z \, \Biggl[
-\tfrac{3817759}{100800}
+\tfrac{7}{4} \, \zeta_2
+\tfrac{67}{12} \, S_2
+\tfrac{6661}{224} \, \text{ln}(m_Z^2)
-\tfrac{3}{16} \, \text{ln}(m_Z^2)^2
\Biggr]
\nonumber \\ &+&h_Z^2 \, \Biggl[
+\tfrac{329665169}{5443200}
+\tfrac{21}{4} \, \zeta_2
-\tfrac{111709}{864} \, S_2
-\tfrac{66001}{1120} \, \text{ln}(m_Z^2)
+\tfrac{159}{32} \, \text{ln}(m_Z^2)^2
\Biggr]
\nonumber \\ &+&h_Z^3 \, \Biggl[
-\tfrac{803801701}{9525600}
-3 \, \zeta_2
+\tfrac{28865}{216} \, S_2
+\tfrac{1256243}{20160} \, \text{ln}(m_Z^2)
-\tfrac{3}{4} \, \text{ln}(m_Z^2)^2
\Biggr]
\nonumber \\ &+&h_Z^4 \, \Biggl[
+\tfrac{275880109013}{4115059200}
+3 \, \zeta_2
-\tfrac{276031}{5832} \, S_2
-\tfrac{2789701}{40320} \, \text{ln}(m_Z^2)
+\tfrac{3}{4} \, \text{ln}(m_Z^2)^2
\Biggr]
\nonumber \\ &+&h_Z^5 \, \Biggl[
-\tfrac{199097721241}{2514758400}
-3 \, \zeta_2
+\tfrac{96601}{1296} \, S_2
+\tfrac{17481691}{221760} \, \text{ln}(m_Z^2)
-\tfrac{3}{4} \, \text{ln}(m_Z^2)^2
\Biggr] \nonumber
\end{eqnarray}
\begin{eqnarray}
R^{\msb}_5=
 & & \, \Biggl[
+\tfrac{1180117}{18144}
+10 \, \zeta_2
-\tfrac{17483}{72} \, S_2
-\tfrac{58129}{1680} \, \text{ln}(m_Z^2)
+3 \, \text{ln}(m_Z^2)^2
\Biggr]
\nonumber \\ &+&h_Z \, \Biggl[
-\tfrac{49494491}{1360800}
+\tfrac{9}{4} \, \zeta_2
+\tfrac{1651}{216} \, S_2
+\tfrac{111187}{3360} \, \text{ln}(m_Z^2)
+\tfrac{1}{8} \, \text{ln}(m_Z^2)^2
\Biggr]
\nonumber \\ &+&h_Z^2 \, \Biggl[
+\tfrac{973643677}{19051200}
+7 \, \zeta_2
-\tfrac{12185}{108} \, S_2
-\tfrac{230051}{3360} \, \text{ln}(m_Z^2)
+\tfrac{97}{16} \, \text{ln}(m_Z^2)^2
\Biggr]
\nonumber \\ &+&h_Z^3 \, \Biggl[
-\tfrac{405345006701}{4115059200}
-4 \, \zeta_2
+\tfrac{1116793}{5832} \, S_2
+\tfrac{195907}{2688} \, \text{ln}(m_Z^2)
-\text{ln}(m_Z^2)^2
\Biggr]
\nonumber \\ &+&h_Z^4 \, \Biggl[
+\tfrac{2758671944087}{45265651200}
+4 \, \zeta_2
-\tfrac{78913}{2916} \, S_2
-\tfrac{6006071}{73920} \, \text{ln}(m_Z^2)
+\text{ln}(m_Z^2)^2
\Biggr]
\nonumber \\ &+&h_Z^5 \, \Biggl[
-\tfrac{33815515952111}{407390860800}
-4 \, \zeta_2
+\tfrac{1210235}{13122} \, S_2
+\tfrac{3432923}{36960} \, \text{ln}(m_Z^2)
-\text{ln}(m_Z^2)^2
\Biggr] \nonumber
\end{eqnarray}

\end{document}